\documentclass[twocol]{ametsocV6.1}

\usepackage{amsmath,amsfonts,amssymb,bm}
\usepackage{mathptmx}
\usepackage{newtxtext}
\usepackage{newtxmath}
\usepackage{nicefrac}

\title{Assessing the Spurious Impacts of Ice-Constraining Methods on the Climate Response to Sea-Ice Loss using an Idealised Aquaplanet GCM}

\authors{Neil T. Lewis\aff{a}\correspondingauthor{Neil T. Lewis, n.t.lewis@exeter.ac.uk}, Mark R. England\aff{a}, James A. Screen\aff{a}, Ruth Geen\aff{b}, Regan Mudhar\aff{a}, \\ William J. M. Seviour\aff{a}, and Stephen I. Thomson\aff{a}}

\affiliation{\aff{a}{Department of Mathematics and Statistics, University of Exeter, UK}\\[5pt]
    \aff{b}{School of Geography, Earth and Environmental Sciences, University of Birmingham, UK}}

\abstract{Coupled climate model simulations designed to isolate the effects of Arctic sea-ice loss often apply artificial heating, either directly to the ice or through modification of the surface albedo, to constrain sea-ice in the absence of other forcings. Recent work has shown that this approach may lead to an overestimation of the climate response to sea-ice loss. In this study, we assess the spurious impacts of ice-constraining methods on the climate of an idealised aquaplanet general circulation model (GCM) with thermodynamic sea-ice. The true effect of sea-ice loss in this model is isolated by inducing ice loss through reduction of the freezing point of water, which does not require additional energy input. We compare results from freezing point modification experiments with experiments where sea-ice loss is induced using traditional ice-constraining methods, and confirm the result of previous work that traditional methods induce spurious additional warming. Furthermore, additional warming leads to an overestimation of the circulation response to sea-ice loss, which involves a weakening of the zonal wind and storm track activity in midlatitudes. Our results suggest that coupled model simulations with constrained sea-ice should be treated with caution, especially in boreal summer, where the true effect of sea-ice loss is weakest but we find the largest spurious response. Given that our results may be sensitive to the simplicity of the model we use, we suggest that devising methods to quantify the spurious effects of ice-constraining methods in more sophisticated models should be an urgent priority for future work.}

\begin{document}

\maketitle

\section{Introduction}

The Arctic is undergoing rapid climate change, characterised by substantial sea-ice loss and polar amplified warming \citep{2016Sci...354..747N,2010Natur.464.1334S}, which has motivated study of the impacts of sea-ice loss on weather and climate \citep{barnes2015impact,2014NatGe...7..627C,2019NatCC..10...20C}. 

{

To tackle this problem, numerous studies have made use of climate model simulations where sea-ice loss is imposed. One approach has been to use  atmosphere-only general circulation models (AGCMs), by comparing simulations where climatological or future sea-ice concentrations (SIC) are prescribed while sea surface temperature (SST) is held fixed (see \citealp{barnes2015impact} for a review). When sea-ice loss occurs, some studies prescribe a fixed `freezing point' SST (e.g., \citealp{2010JCli...23..333D}), while others prescribe SSTs from simulations of future climate change, to account for sea surface warming associated with sea-ice loss (e.g., \citealp{2013JCli...26.1230S}). \citet{2018JCli...31.9193B} show that the response to sea-ice loss is essentially insensitive to the chosen ice-free SST (see their Figure 4b).  AGCM experiments reveal that while Arctic sea-ice loss is greatest in summer and autumn, the circulation response is greatest in winter \citep{2010JCli...23..333D}. They consistently find that Arctic sea-ice loss induces a weakening of the midlatitude westerlies (see \citealp{2022NatCo..13..727S}, which presents results from the Polar Amplification Model Intercomparison Project, PAMIP; \citealp{2019GMD....12.1139S}), but disagree on other aspects of change caused by sea-ice loss, for example the response of the stratosphere \citep{2018NatGe..11..155S,2022NatCo..13..727S}.

More recently, research into the impact of sea-ice loss on climate has expanded to make use of coupled atmosphere--ocean general circulation models (AOGCMs), which explicitly simulate interactions between the atmosphere, ocean, sea-ice, and land \citep[e.g.,][]{2015JCli...28.2168D,2016JCli...29..401B,2017JCli...30.4547S,2017GeoRL..44.7955M,2019GMD....12.1139S,2020GeoRL..4785788S}. These studies investigate the response of AOGCMs to perturbed sea-ice, by constraining sea-ice area or volume, or both, to match that obtained in simulations of future climate change (using a variety of methods, discussed below), while keeping other climate forcings constant. Comparing AGCM and AOGCM experiments, \citet{2015JCli...28.2168D} find that including ocean coupling results in a warming response that extends to lower latitudes and higher altitudes, and an increase of the Northern Hemisphere zonal wind response by approximately 30\%. Moreover, \citet{2016JCli...29.6841T}, \citet{2018GeoRL..45.4264W} and \citet{2020NatGe..13..275E} show that the remote effects of sea-ice loss on the tropics may depend on the response of the ocean circulation (see also discussion in \citealp{2018NatGe..11..155S}). \citet{2022JCli...35.4665A} investigate the impacts of Antarctic sea-ice loss on climate, and, similarly to \citet{2015JCli...28.2168D}, find that the circulation response is larger in coupled experiments versus uncoupled experiments. \citet{2018NatGe..11..155S} suggest that the circulation response to sea-ice loss appears to be more consistent between different AOGCMs when compared with AGCMs.

}

Sea-ice loss itself is a response to warming due to greenhouse gas (GHG) emissions. Therefore, coupled model studies which seek to isolate the impacts of sea-ice loss, absent other climate forcings, require an additional, artificial energy input to be included in the models, to melt the sea-ice. Various methods to constrain sea-ice have been utilised, including surface albedo reduction \citep{2015JCli...28.2168D,2016JCli...29..401B}, and the `nudging' and `ghost flux' methodologies, which directly add heat to the sea-ice module  \citep{2015JCli...28.2168D,2016JCli...29.6841T,2017JCli...30.4547S,2017GeoRL..44.7955M,2020NatGe..13..275E,2021JCli...34.3751P}. Comparison studies have shown that these methods produce results that are broadly consistent with one another \citep{2018NatGe..11..155S,2020GeoRL..4785788S}. However, \citet{2022JCli...35.3801E} argue that they also share a common, `spurious' side effect; namely that the surface temperature response to sea-ice loss is overestimated, as the direct warming response to artificial energy input (required to melt the ice) is erroneously included as a response to sea-ice loss, when in reality it is the cause.

{
Support for \citet{2022JCli...35.3801E}'s argument has been offered by \citet{2024ERCli...3a5003F}, who develop a technique to correct the response obtained in AOGCMs with constrained sea-ice post-hoc, using multi-parameter pattern scaling. They build on \citet{2017JCli...30.2163B}, who decompose the response of some field $Z$ as: \begin{equation}
\delta Z = \left.\frac{\partial Z}{\partial T_{l}}\right\rvert_{I} \delta T_{l} + \left.\frac{\partial Z}{\partial I}\right\rvert_{T_{l}}\delta I, 
\end{equation}
where $T_{l}$ is low-latitude SST and $I$ is sea-ice area. The first term represents the response of $Z$ to a change in $T_{l}$, absent a change in $I$, and the second term represents the response of $Z$ to a change in $I$, absent a change in $T_{l}$ (i.e., the response that scales with sea-ice loss). This equation can be solved for the sensitivities $(\partial Z/\partial T_{l})\rvert_{I}$ and $(\partial Z/\partial I)\rvert_{T_{l}}$ given output from \emph{two} pairs of experiments; for example, a control experiment, compared against a simulation of climate change with increased GHG emissions, and a simulation where sea-ice is constrained through application of artificial heating to the Arctic. In light of \citet{2022JCli...35.3801E}'s result that artificial heating drives a spurious warming response, \citet{2024ERCli...3a5003F} propose that this approach can be adapted to determine the \emph{true effect} of sea-ice loss, which is the component of the response driven by energy input due to changes in the surface albedo associated with reduced sea-ice coverage. This is achieved by replacing the scaling variable $I$ with one that accounts for the spurious forcing used to induce sea-ice loss (in addition to the ice-loss itself). 

Considering AOGCM experiments forced by (i) increased GHG emissions and (ii) albedo modification, \citet{2024ERCli...3a5003F} replace $I$ with the change in net all-sky TOA shortwave north of $70^{\circ}\,\text{N}$, $\text{SW}_{\text{TOA}}$, which accounts for increased energy input due to reduced sea-ice coverage (which reduces the surface albedo), as well as spurious energy input in the albedo modification experiment (due to the fact the ice is artificially darker). Doing so, they find the annual-mean high-latitude warming attributable to sea-ice loss is reduced from roughly $8\,\text{K}$ when $I$ is used to just over $5\,\text{K}$ when $\text{SW}_{\text{TOA}}$ is used (compared to a total response of roughly $8\,\text{K}$ to increased GHG emissions), and the magnitude of the midlatitude zonal wind response is reduced by roughly $50\%$ (while retaining its spatial structure). Noting that \citet{2015JCli...28.2168D} find ocean coupling enhances the midlatitude jet response by roughly 30\%, relative to the response in an AGCM, \citeauthor{2024ERCli...3a5003F}'s results imply that the jet response to Arctic sea-ice loss could actually be damped, rather than enhanced, by ocean coupling. It is worth noting that artificial heating does not necessarily strengthen all aspects of the climate response to sea-ice loss. For example, \citet{2023...NL} use idealised general circulation model (GCM) experiments to show that nudging may artificially weaken the response of Arctic surface temperature persistence to sea-ice loss. }

The analysis presented in \citet{2022JCli...35.3801E} is based on simulations run using a dry, diffusive energy balance model (EBM). This choice allows \citeauthor{2022JCli...35.3801E} to compare the surface temperature response to various ice-constraining methods against the true effect of sea-ice loss on temperature, which, due to the EBM's simplicity, can be determined analytically. However, the dry EBM precludes them from assessing the extent to which the spurious, additional warming response is accompanied by artefacts in the response of atmospheric circulation to sea-ice loss (which is not represented in the EBM). 
Additionally, it omits processes important to the real climate system, including the poleward transport of latent heat by water vapour (which can drive polar amplified warming in the absence of sea-ice loss; e.g.,  \citealp{2018JCli...31.5811M,2021GeoRL..4894130F}), as well as feedbacks associated with clouds \citep{england2024robust}, and the response of poleward heat transport to sea-ice loss due to changes in atmospheric circulation \citep{2011GeoRL..3817704H}. \citet{2024ERCli...3a5003F} extend the analysis of \citet{2022JCli...35.3801E} to a moist EBM, and show that their conclusions are unaltered by this addition. {
In addition, their pattern scaling approach offers a route to identifying the impacts of spurious warming on the circulation response to sea-ice loss in AOGCMs. However, it should be noted that pattern-scaling only approximates the true effect of sea-ice loss (compared with an exact quantification in the EBM framework). In addition, \citet{2024ERCli...3a5003F} have less success correcting for artificial heat in experiments that use a ghost flux to constrain sea-ice, as in this scenario the appropriate choice for the replacement scaling variable is less clear.}

The objective of this study is to extend the work of \citet{2022JCli...35.3801E} and \citet{2024ERCli...3a5003F} using the \texttt{Isca} idealised GCM framework \citep{2018GMD....11..843V}, configured as an aquaplanet with a slab ocean and thermodynamic sea ice \citep[similar to that described in][]{2021GeoRL..4894130F,chung2023sea}. The true effect of sea-ice loss on the climate of this model (i.e., the effect of sea-ice loss, absent the effect of additional heating required to melt ice) can be obtained by decreasing the freezing point of water to reduce ice coverage. This method does not require surface warming in order to induce sea-ice loss. Instead, sub-freezing open ocean can exist, and all additional energy input to the climate system arises due to changes in the surface albedo (which occur due to sea-ice loss, as opposed to artificial modification of the ice albedo). By comparing the climate response to sea-ice loss induced by (i) albedo modification and (ii) a simplified ice-nudging methodology, with that induced by freezing point modification, we are able to isolate the spurious side effects of methodologies (i) and (ii) on surface temperature and large-scale atmospheric circulation in the idealised GCM. By conducting this study, we aim to investigate the extent to which the apparently larger response in AOGCMs compared to AGCMs can be described as an artefact of the method used to constrain sea-ice. The remainder of this paper is structured as follows. Section \ref{sec:gcm} presents a description of the idealised GCM we use, and Section \ref{sec:expts} details our experiment design. Our results are presented in Section \ref{sec:results}. Finally, discussion and a summary of our main conclusions are included in Section \ref{sec:discuss}.

\section{Model Description}\label{sec:gcm}


We run numerical experiments using \texttt{Isca}, a framework for modeling the atmospheres of the Earth and other planets at varying levels of complexity \citep{2018GMD....11..843V}. The model used for this study constitutes an idealised aquaplanet GCM, configured with a semi-grey radiative transfer scheme, including seasonally varying insolation, and a heavily simplified representation of moist processes that omits clouds entirely \citep[following][]{2007JAtS...64.1959F,2008JCli...21.3815O}. The surface is represented as a slab ocean with prescribed ocean heat transport \citep[following][]{2013JCli...26..754M}, and features a simple thermodynamic sea-ice code based on \citet{2022JAMES..1402671Z}. This configuration is similar to that used by other studies that investigate the climate response to sea-ice loss with an idealised model \citep{2021GeoRL..4894130F,2022JCli...35.2633S,chung2023sea,2023...NL}. 

\subsection{Surface energy budget}

For ice-free conditions, the model's surface energy budget evolves according to \begin{align}
C\frac{\partial T_{\text{ml}}}{\partial t}&=-F_{\text{atm}} + \nabla\cdot\mathbf{F}_{\text{ocean}}, \\ 
T_{\text{s}}&=T_{\text{ml}},
\end{align}
where $T_{\text{ml}}$ is the ocean mixed-layer temperature, $T_{\text{s}}$ is the surface temperature, $F_{\text{atm}}$ denotes the net downward radiative and turbulent surface heat flux, and $\nabla\cdot\mathbf{F}_{\text{ocean}}$ represents prescribed poleward ocean heat transport. $C=\rho_{\text{w}}c_{\text{w}}d$ is the heat capacity of the mixed-layer ocean, where $\rho_{\text{w}}=1035\,\text{kg}\,\text{m}^{-3}$ is the density of sea water, $c_{\text{w}}=3989\,\text{J}\,\text{kg}^{-1}\,\text{K}^{-1}$ is the specific heat capacity of water, and $d=30\,\text{m}$ is the depth of the mixed-layer (chosen to obtain a seasonal cycle with a realistic ampltiude and lag). 

When the surface temperature drops below the freezing temperature $T_{\text{freeze}}$, sea-ice is allowed to grow, and the surface energy budget is given by \begin{align}
C\frac{\partial T_{\text{ml}}}{\partial t}&=-F_{\text{base}} + \nabla\cdot\mathbf{F}_{\text{ocean}}, \\ 
L\frac{\partial h}{\partial t} &= F_{\text{atm}} - F_{\text{base}}, \label{eq:thick} \\ 
F_{\text{base}} &= F_{0}\left(T_{\text{ml}} - T_{\text{freeze}}\right), \\ 
F_{\text{atm}} &= F_{i} \equiv k_{i}\frac{T_{\text{freeze}}-T_{\text{s}}}{h}. \label{eq:cond}
\end{align}
This representation of thermodynamic sea-ice is based on the `zero layer' model proposed by \citet{1976JPO.....6..379S}, and our implementation exactly follows that described by \citet{2022JAMES..1402671Z}. Above, $h$ is the sea-ice thickness, $F_{\text{base}}$ is the basal heat flux from the mixed-layer into the ice, which linearly depends on the difference between $T_{\text{ml}}$ and the temperature at the ice base (the melting temperature, which for simplicity we set equal to $T_{\text{freeze}}$). When ice is present, the surface temperature is determined implicitly via a balance between $F_{\text{atm}}$ and $F_{i}$, the conductive heat flux through the ice, unless this procedure yields $T_{\text{s}}>T_{\text{freeze}}$, in which case Equation \ref{eq:cond} is replaced with $T_{\text{s}}=T_{\text{freeze}}$ (representing surface melt). Above, the coefficient $F_0$ is set to $120\,\text{W}\,\text{m}^{-2}\,\text{K}^{-1}$, the thermal conductivity of ice is set to $k_{i}=2\,\text{W}\,\text{m}^{-1}\,\text{K}^{-1}$, and the latent heat of fusion is $L=3\times10^{8}\,\text{J}\,\text{m}^{-3}$. For our control simulation, we set $T_{\text{freeze}}=-2^{\circ}\,\text{C}$ (roughly the freezing point of salt water), to ensure a realistic latitude for the ice-edge. 

Finally, a semi-realistic, time-invariant representation of ocean heat transport is included in the model, using the functional form proposed by \citet{2013JCli...26..754M}: \begin{equation}
\nabla\cdot\mathbf{F}_{\text{ocean}} = \frac{Q_0}{\cos\vartheta}\left(1-\frac{2\vartheta^{2}}{\vartheta_0^{2}}\right)\exp\left(-\frac{\vartheta^{2}}{\vartheta_0^{2}}\right), \end{equation}
where $\vartheta$ is latitude, $\vartheta_0=\pm16^{\circ}$, and we set the amplitude to be $Q_0=30\,\text{W}\,\text{m}^{-2}$.

\subsection{Radiative transfer}

Radiative transfer is represented using a simplified semi-grey scheme with fixed optical depths \citep[similar to][]{2006JAtS...63.2548F,2007JAtS...64.1959F,2008JCli...21.3815O}. For longwave (infrared) radiation, upward and downward fluxes are computed using the two-stream approximation: \begin{align}
\frac{\text{d}F^{\uparrow}_{\text{lw}}}{\text{d}\tau_{\text{lw}}} &= F^{\uparrow}_{\text{lw}} - \sigma_{\text{sb}} T^4, \\ 
\frac{\text{d}F^{\downarrow}_{\text{lw}}}{\text{d}\tau_{\text{lw}}} &= \sigma_{\text{sb}} T^4 -F^{\downarrow}_{\text{lw}},
\end{align}
where $\sigma_{\text{sb}}=5.67\times10^{-8}\,\text{W}\,\text{m}^{-2}\,\text{K}^{-4}$ is the Stefan--Boltzmann constant, $T$ is temperature, and $\tau_{\text{lw}}$ is the longwave optical depth, defined by the function \begin{equation}
\tau_{\text{lw}} \equiv \left[f\sigma+\left(1-f\right)\sigma^{4}\right]\left[\tau_{\text{e}}+\left(\tau_{\text{p}}-\tau_{\text{e}}\right)\sin^{2}\vartheta\right], 
\end{equation}
where $f=0.2$, $\sigma=p/p_{\text{s}}$ is pressure normalized by the surface pressure, $\tau_{\text{e}}$ is the longwave optical depth at the equator, and $\tau_{\text{p}}$ is the longwave optical depth at the pole. For our control experiment, we set $\tau_{\text{e}}=7.2$ and $\tau_{\text{p}}=3.6$. At the lower boundary, $F^{\uparrow}_{\text{lw, sfc}} = \sigma_{\text{sb}}T_{\text{s}}^{4}$, and at the top-of-atmosphere, $F^{\downarrow}_{\text{lw, TOA}} = 0$. 

For shortwave (visible light) radiation, the downward flux is given by \begin{equation}
F^{\downarrow}_{\text{sw}} = \left(1-\alpha_{\text{TOA}}\right)S_{\text{TOA}}\exp\left(-\tau_{\text{sw}}\sigma^{2}\right), 
\end{equation}
where $\tau_{\text{sw}}=0.22$ is the shortwave optical depth, and $1-\alpha_{\text{TOA}}\equiv\left[0.75 + 0.15 \times P_{2}(\sin\vartheta)\right]$ is a latitudinally varying co-albedo, which is included to account for the missing effect of clouds. The top-of-atmosphere insolation $S_{\text{TOA}}$ is computed for a circular orbit and the Earth's obliquity, excluding the diurnal cycle, and assuming a solar constant of $S_0=1360\,\text{W}\,\text{m}^{-2}$. All shortwave radiation reflected at the surface is assumed to immediately escape to space, so that $F^{\uparrow}_{\text{sw}}\left(\tau_{\text{sw}}\right) = \alpha_{\text{sfc}}F^{\downarrow}_{\text{sw, sfc}}$, where $\alpha_{\text{sfc}}$  is the surface albedo. Open ocean has an albedo of $\alpha_{\text{ocean}}=0.1$, and for our control simulation we set the albedo of sea-ice to $\alpha_{\text{ice}}=0.55$. 

\subsection{Sub grid-scale processes}

Simplified representations of sub grid-scale processes are included in the model, exactly following \citet{2008JCli...21.3815O}. Convection is parameterised using the `Simple Betts--Miller' scheme of \citet{2007JAtS...64.1959F}, incorporating the modifications to shallow convection implemented by \citet{2008JCli...21.3815O}. A grid scale condensation scheme is included to adjust humidity and temperature whenever there is large-scale saturation of a gridbox (i.e., relative humidity exceeding 100\%). Surface fluxes of sensible and latent heat are computed using standard bulk aerodynamic formulae (Equations 9--11 in \citealp{2006JAtS...63.2548F}), and boundary layer turbulence is parametrised using a $k$-profile scheme similar to \citet{1986BoLMe..37..129T}. Diffusion coefficients are obtained from Monin--Obukhov similarity theory, using the implementation described by \citet{2008JCli...21.3815O}. 

\subsection{Dynamical core}

\texttt{Isca} uses the Geophysical Fluid Dynamics Laboratory spectral dynamical core to integrate the primitive equations forwards in time. For the present study, we configure the model with a T42 spectral resolution, corresponding to a latitude--longitude resolution of roughly $2.5^{\circ}$, and a timestep of $900\,\text{s}$. In the vertical, there are 30 layers, distributed according to $\sigma=\exp\left[-5\left(0.05\tilde{z}+0.95\tilde{z}^{3}\right)\right]$, where $\tilde{z}$ is evenly spaced on the unit interval. 

\section{Experiment design}\label{sec:expts}

\subsection{Control experiment and `climate change' experiments.}
Table \ref{t1} summarises the various experiments we run using the idealised GCM. For our control simulation, denoted CTRL, we run the model using the parameters defined in the previous subsections for 50 years, starting from an isothermal, quiescent initial condition. The final 20 years of this simulation are used for analysis. In addition, we run three `climate change' experiments, where the longwave optical depth is increased by a multiplicative pre-factor: $\tau_{\text{lw}}=\tau_{\text{lw, ctrl}}\times\text{ODP}$. We consider three values, $\text{ODP}=1.05$, $1.1$, and $1.2$, which yield global mean surface temperature increases of $\Delta T_{\text{s}}=1.5$, $2.8$, and $4.4\,\text{K}$, respectively. These experiments are denoted ODP$X$, where $X$ indicates the value of ODP used. Each ODP experiment is run for 50 years, using the start of the $31^{\text{st}}$ year of the CTRL simulation as an initial condition. The final 20 years of each ODP experiment are used for analysis.  

\subsection{Sea-ice loss experiments}\label{sec:ice_expts}

\begin{table*}[t]
\caption{Summary description of idealised aquaplanet GCM experiments.}\label{t1}

\begin{center}
\vspace*{-.1in}
{\renewcommand{\arraystretch}{1.5}
\begin{tabular}{lll}
\hline\hline
Experiment & Intervention & Notes\\
\hline
CTRL & None & Model configuration described in Section \ref{sec:gcm}. \\
ODP & Longwave optical depth increase & \parbox[t]{4in}{Longwave optical depth is increased by a multiplicative pre-factor so that $\tau_{\text{lw}} = \tau_{\text{lw,ctrl}}\times\text{ODP}$. We consider three values, $\text{ODP}=1.05$, $1.1$, and $1.2$. Experiments run with these values are denoted ODP1.05, ODP1.1, and ODP1.2, respectively.}\phantom{.}  \\
FRZ & Freezing point modification & \parbox[t]{4in}{Sea-ice loss induced by reducing the freezing point of water, $T_{\text{frz}}$. This method does not require additional energy input to melt ice. Experiments are run for a wide range of $T_{\text{freeze}}$ (see Section \ref{sec:expts}). The experiments with $T_{\text{freeze}}=-3.75$, $-5$, and $-10^{\circ}\,\text{C}$, denoted FRZ3.75, FRZ5, and FRZ10, respectively, yield annually-averaged SIAs that best mach those of ODP1.05, ODP1.1, and ODP1.2, respectively.}   \\ 
NDG\_NC & Nudging & \parbox[t]{4in}{Sea-ice loss induced by applying a heating to the sea-ice at  latitudes and times where it is not present in the target climate (see Equation \ref{eq:Fnudge}). NDG experiments target the SIA obtained in the ODP experiments, and are denoted NDG\_NC1.05, NDG\_NC1.1, and NDG\_NC1.2.}  \\ 
NDG & Nudging & \parbox[t]{4in}{As above, but with a corrective cooling applied in regions where sea-ice loss does not occur, so that the globally-averaged energy input due to nudging is zero. Three experiments are run: NDG1.05, NDG1.1, and NDG1.2.} \\ 
ALB & Albedo modification & \parbox[t]{4in}{Sea-ice loss induced by reducing the ice albedo. Experiments are run for a wide range of $\alpha_{\text{ice}}$ (see Section \ref{sec:expts}). Experiments with $\alpha_{\text{ice}}=0.45$ and $0.1$, denoted ALB.45 and ALB.1, obtain annually-averaged SIAs that best match those in ODP1.05 and ODP1.1, respectively. Experiments with $\alpha_{\text{ice}}=0.48$ and $0.35$, denoted ALB.48 and ALB.35, obtain JJA SIAs that best match those in ODP1.05 and ODP1.1. We were unable to replicate te degree of sea-ice loss obtained in ODP1.2 using any value of $\alpha_{\text{ice}}\geq0.1$ ($\alpha_{\text{ocean}}=0.1$ is the value for the surface albedo of ice-free ocean).} \\ \\[-10pt]
\hline
\end{tabular}
}
\end{center}
\end{table*}

\begin{figure}[!t]
    \centering 
    \includegraphics[width=.48\textwidth]{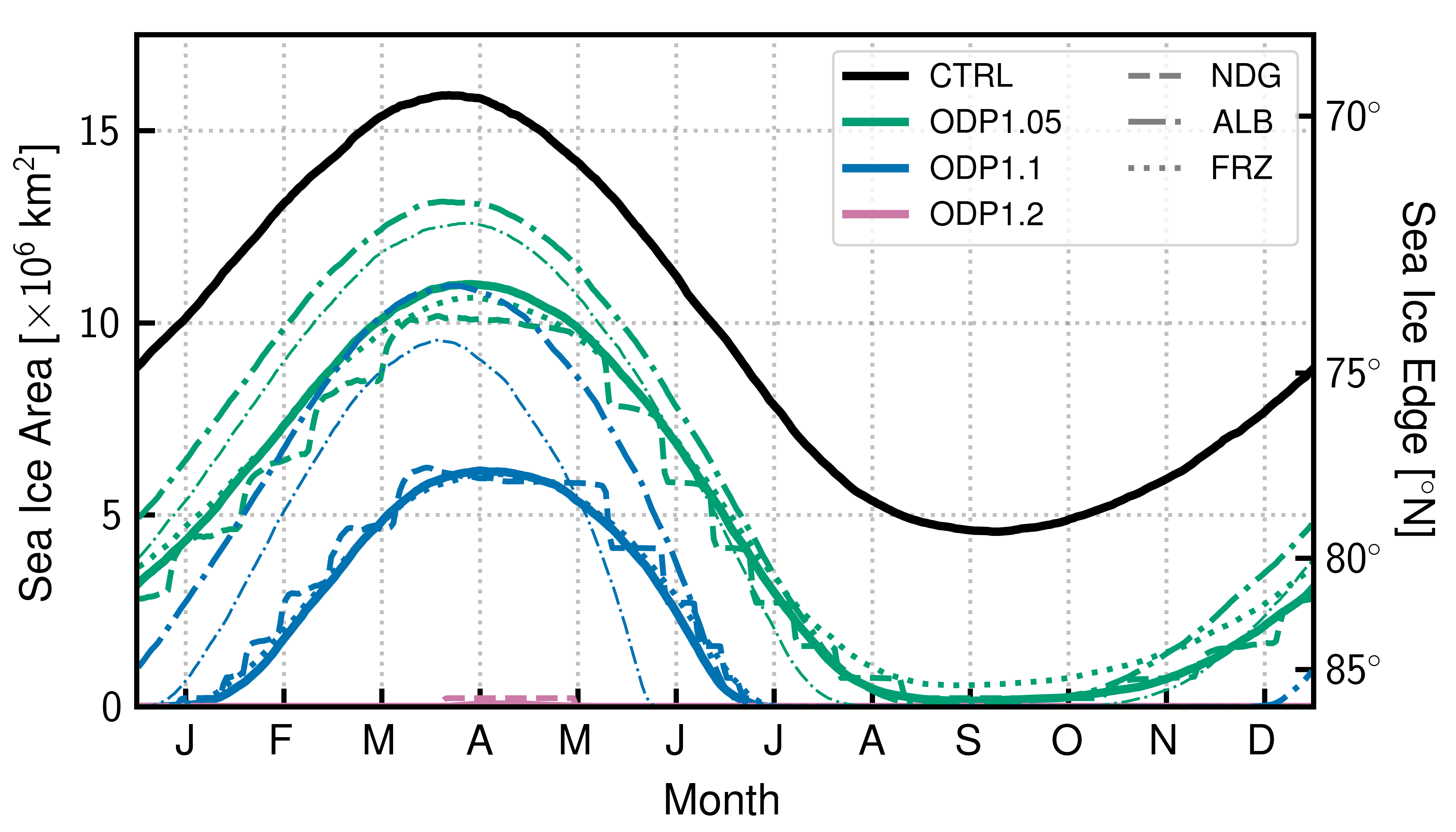}
    \caption{Seasonal cycle of Northern Hemisphere sea-ice area (SIA) obtained in idealised GCM experiments. Solid lines show the CTRL experiment and ODP experiments, and broken lines show experiments with SIA constrained to match that obtained in the ODP experiments. Colour indicates correspondence between ODP and ice-constrained simulations. Dashed lines show SIA in NDG experiments, dash-dot lines show SIA in FRZ experiments (FRZ3.75, FRZ5, and FRZ10), and dotted lines show SIA in ALB experiments. For ALB experiments, thicker dotted lines show experiments that best match the summer (JJA) SIA in the corresponding ODP runs (ALB.48 and ALB.35), and thinner lines show experiments that best match the annually-averaged SIA (ALB.45 and ALB.1).}\label{fig:SIA_seasons}
\end{figure}

To investigate the impacts of sea-ice loss on climate, we run a suite of `counterfactual' experiments with constrained sea-ice, absent other climate forcings. We consider experiments with a modified freezing temperature for ice, which capture the true effect of sea-ice loss on the model climate, alongside a simplified implementation of nudging, and albedo modification (targeting both summer SIA and annually-averaged SIA), which are commonly used methodologies for constraining sea-ice in AOGCMs. Each approach is described in detail in the subsections that follow. All sea-ice loss experiments use the start of the $31^{\text{st}}$ year of the CTRL simulation as an initial condition, and are run for 50 years, with the final 20 years used for analysis.  Seasonally varying, zonally-averaged SIA is shown in Figure \ref{fig:SIA_seasons} for selected experiments with constrained sea-ice, compared against the CTRL and ODP runs. 

\subsubsection{Freezing point modification}

Our choice to configure \texttt{Isca} with a slab ocean and thermodynamic sea-ice means that we can induce sea-ice loss by reducing the freezing temperature in the thermodynamic sea-ice code. This method allows us to vary the sea-ice area without directly inducing surface warming. Instead, all warming that arises is due to changes in the sea-ice albedo (reduced in regions which transition from an ice covered state to an ice free state), which we define as the true effect of sea-ice loss (consistent with \citealt{2022JCli...35.3801E} and \citealt{2024ERCli...3a5003F}). We have checked this by configuring the model so that the albedos of open ocean and sea-ice are the same; in this configuration, the annually- and zonally-averaged surface temperature is insensitive to $T_{\text{freeze}}$. We note that it would be difficult to implement this methodology using a more sophisticated model, featuring a dynamic ocean and dynamic sea-ice, as it would require an artificial extrapolation of the equation of state for water to sub-freezing temperatures.

We have run experiments using the following values for the freezing point: $T_{\text{freeze}}=-2.5$, $-3$, $-3.5$,  $-3.75$, $-4$, $-5$, $-6$, and $-10\,^{\circ}\,\text{C}$. In the CTRL configuration, $T_{\text{freeze}}=-2^{\circ}$, to ensure a realistic value for the ice-edge; this is our reason for beginning this sequence of counterfactual experiments with $T_{\text{freeze}}=-2.5\,^{\circ}\,\text{C}$. We label these experiments FRZ$X$, where $X$ indicates the magnitude $\lvert T_{\text{freeze}}\rvert$ used. The experiments FRZ$3.75$, FRZ$5$, and FRZ$10$ yield annually-averaged SIAs that best match those of the ODP$1.05$, ODP$1.1$, and ODP$1.2$ experiments, respectively. Seasonally varying SIA obtained from the FRZ$3.75$, FRZ$5$, and FRZ$10$ experiments is shown using dotted lines in Figure \ref{fig:SIA_seasons}. This figure demonstrates that these experiments accurately capture the seasonal cycle of sea-ice loss in the ODP experiments, in addition to recovering the correct annual-mean value.

\subsubsection{Albedo modification}

For our albedo modification experiments, we vary the sea-ice albedo, $\alpha_{\text{ice}}$. We consider the following values: $\alpha_{\text{ice}}=0.5$, $0.485$, $0.48$, $0.475$, $0.45$, $0.4$, $0.35$, $0.3$, $0.2$, $0.15$, and $0.1$ (i.e., the same value as $\alpha_{\text{ocean}}$). These experiments are denoted ALB$X$, where $X$ indicates the value of $\alpha_{\text{ice}}$ used.

Reducing the sea-ice albedo has a limited effect during the polar night, and is therefore most effective at reducing summer SIA. This means that the seasonal cycle of sea-ice loss under albedo modification is skewed towards summer months, and a choice must be made regarding whether to target the summer sea-ice loss or annually-averaged sea-ice loss due to climate change \citep{2016JCli...29..401B,2020GeoRL..4785788S,2022JCli...35.3801E}. Through our parameter sweep over $\alpha_{\text{ice}}$, we cover both options. The ALB$.45$ and ALB$.1$ experiments obtain the annually averaged SIA that best matches that in the  ODP$1.05$ and ODP$1.1$ experiments, respectively. The seasonal cycle of SIA in ALB$.45$ and ALB$.1$ is presented in Figure \ref{fig:SIA_seasons} as thin dash-dot lines, showing how targeting annually-averaged SIA leads to excessive sea-ice loss in summer, and restricted sea-ice loss in winter (compared with the corresponding ODP experiments). ALB$.48$ and ALB$.35$ yield the best match to the summer sea-ice loss obtained in ODP$1.05$ and ODP$1.1$, respectively. These experiments are shown in Figure \ref{fig:SIA_seasons} as thick dash-dot lines, revealing a much better match to the onset of summer sea-ice loss, at the expense of more severely under-representing sea-ice loss in winter. We note that we were unable to replicate the degree of sea-ice loss obtained in the  ODP$1.2$ experiment using any of the values for $\alpha_{\text{ice}}$ listed above.

\subsubsection{Nudging}\label{sec:ndg_methods}

Finally, we run the model in the CTRL configuration, but with the sea-ice area nudged towards that obtained in each of the ODP experiments described above.

Nudging is implemented by adding an additional term to the equation for sea-ice thickness evolution (Equation \ref{eq:thick}): \begin{equation} 
L\frac{\partial h}{\partial t} = \cdots + F_{\text{nudge}},  \label{eq:nudge}
\end{equation}
which is non-zero at latitudes and times when sea-ice is present but $h_{\text{target}}=0$, where $h_{\text{target}}$ is the target sea-ice distribution (a function of latitude and day-of-year, obtained from one of the ODP experiments described above). In the present study, we use a simple implementation of $F_{\text{nudge}}$ following \citet{2022JCli...35.3801E}, given by \begin{equation}
F_{\text{nudge\_nc}}=\begin{cases} -Lh/\tau & \text{when}\quad h_{\text{target}}=0, \\ 0 & \text{otherwise,} \end{cases} \label{eq:Fnudge}
\end{equation}
where $\tau=86400\,\text{s}$ is chosen for the nudging timescale. This approach adds energy to the system; to correct for this, we include an additional, constant correction term, $F_{\text{correct}}$. This term is included at grid points where $F_{\text{nudge\_nc}}=0$, with the magnitude of $F_{\text{correct}}$ set to that which ensures the global, area-weighted average of
\begin{equation} F_{\text{nudge}}=F_{\text{nudge\_nc}}+F_{\text{correct}}
\end{equation} is zero ($F_{\text{correct}}$ is computed at each timestep). We note that while this approach achieves no net energy input from the nudging process, it does so by introducing an unphysical cooling effect to low latitudes.

Experiments including the correction described above are referred to using the name NDG$X$, and experiments with no correction are referred to using NDG\_NC$X$, where $X$ indicates the value of ODP used in the simulation from which $h_{\text{target}}$ is derived. The seasonal cycle of SIA obtained from the NDG$X$ experiments is shown with dashed lines in Figure \ref{fig:SIA_seasons}, demonstrating that the simple nudging implementation adequately constrains SIA to match each ODP experiment. The seasonal cycle of SIA in the NDG\_NC$X$ experiments is essentially identical to that in the NDG$X$ runs.

\section{Results}\label{sec:results}

\subsection{True effect of sea-ice loss on idealised model climate}\label{sec:true_effect}

\begin{figure*}[!t]
    \centering 
    \includegraphics[width=.975\textwidth]{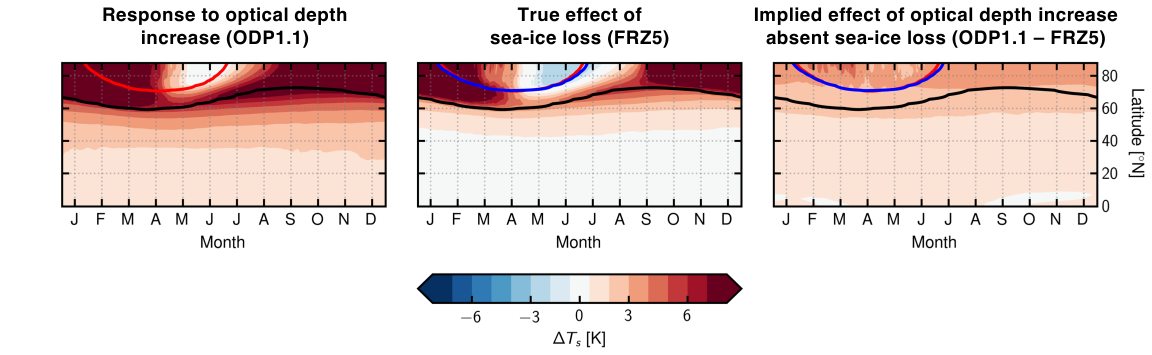}
    \caption{Seasonal surface temperature response obtained in the ODP1.1 and FRZ5 experiments (which obtains a SIA that matches that of ODP1.1; see Figure \ref{fig:SIA_seasons}). Left panel: difference between ODP1.1 and CTRL experiments. Centre panel: difference between FRZ5 and CTRL experiments, which represents the true contribution of sea-ice loss to the response in ODP1.1. Right panel: difference between ODP1.1 and FRZ5, which represents the implied effect of increasing optical depth, minus the effect of sea-ice loss. Red, blue, and black lines represent the sea-ice edge in the ODP1.1, FRZ5, and CTRL experiments, respectively.}\label{fig:true_effect}
\end{figure*}

We begin by describing the true effect of sea-ice loss on surface temperature and atmospheric circulation in the idealised GCM, using the ODP$1.1$ and FRZ$5$ experiments as an illustrative example. Figure \ref{fig:true_effect} shows the seasonally varying response of zonally-averaged surface temperature obtained with each experiment, relative to CTRL, as well as their difference (ODP$1.1-$FRZ$5$). 
In ODP$1.1$, the surface temperature response is polar amplified, with $\Delta T_{\text{s}}$ exceeding $8\,\text{K}$ in polar regions ($>70^{\circ}\,\text{N}$) for much of the year, compared with more modest warming of roughly $3\,\text{K}$ in midlatitudes and $1.5\,\text{K}$ in the tropics. In the Arctic, the surface temperature response is suppressed in late spring and early summer (May through July) compared with the rest of the year, whereas at lower latitudes there is far less seasonal variation. This seasonal cycle of polar amplification is consistent with that obtained in more sophisticated climate models \citep{2009GeoRL..3616704L,2021FrEaS...9..725H} and identified in observations \citep{2010GeoRL..3716707S}.

Comparison between the ODP$1.1$ and FRZ$5$ experiments shows that in our idealised GCM, the majority of polar warming in ODP1.1, as well as its seasonality, is  attributable to sea-ice loss. However, it is important to point out that the residual warming (ODP$1.1-$FRZ$5$), which quantifies the implied effect of increasing optical depth, absent the effects of sea-ice loss, is still polar amplified, consistent with previous work using idealised models (\citealt{2018JCli...31.5811M}; \citealt{2021GeoRL..4894130F}). In the tropics, the residual warming is comparable to the total warming in ODP1.1, which indicates that it is primarily a direct response to increasing optical depth, as opposed to being an effect of sea-ice loss.

\begin{figure}[!t]
    \centering 
    \includegraphics[width=.475\textwidth]{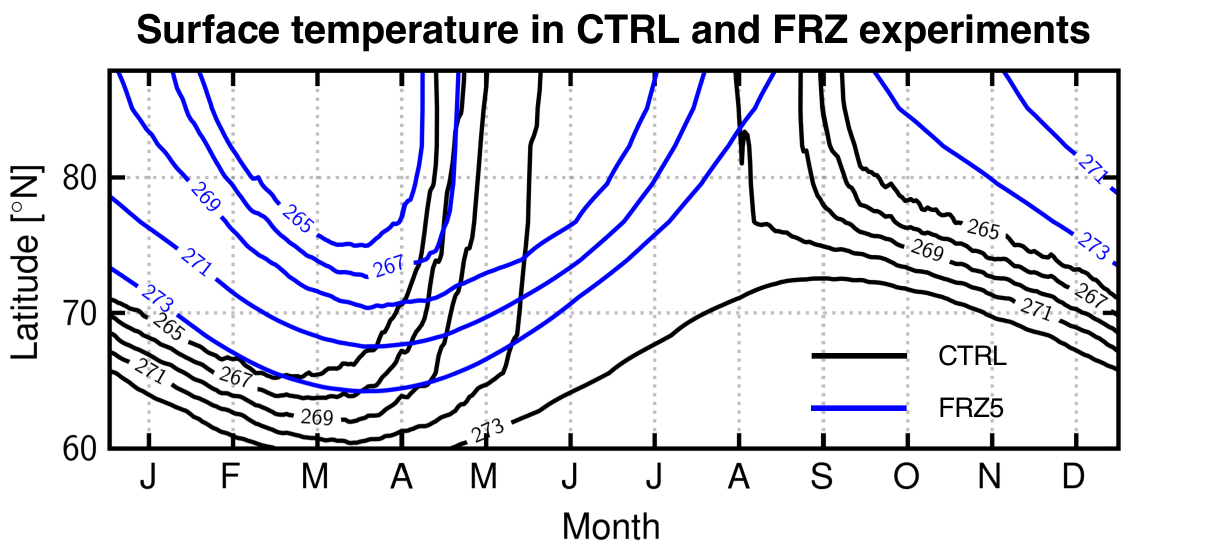}
    \caption{Surface temperature obtained in the CTRL (black contours) and FRZ5 (blue contours) experiments. The freezing temperature for CTRL is $T_{\text{freeze}}=271.15\,\text{K}$, and for FRZ5 is $T_{\text{freeze}}=268.15\,\text{K}$. Surface temperature increase in summer is slowed by the latent heating required to melt sea-ice. Sea-ice retreat occurs earlier in the year for the FRZ5 sea-ice loss experiment compared with CTRL. This causes surface temperature in FRZ5 to be briefly cooler than in CTRL (between May and July; see Figure \ref{fig:true_effect}, centre panel). }\label{fig:true_effect2}
\end{figure}

\begin{figure*}[!t]
    \centering 
    \includegraphics[width=\textwidth]{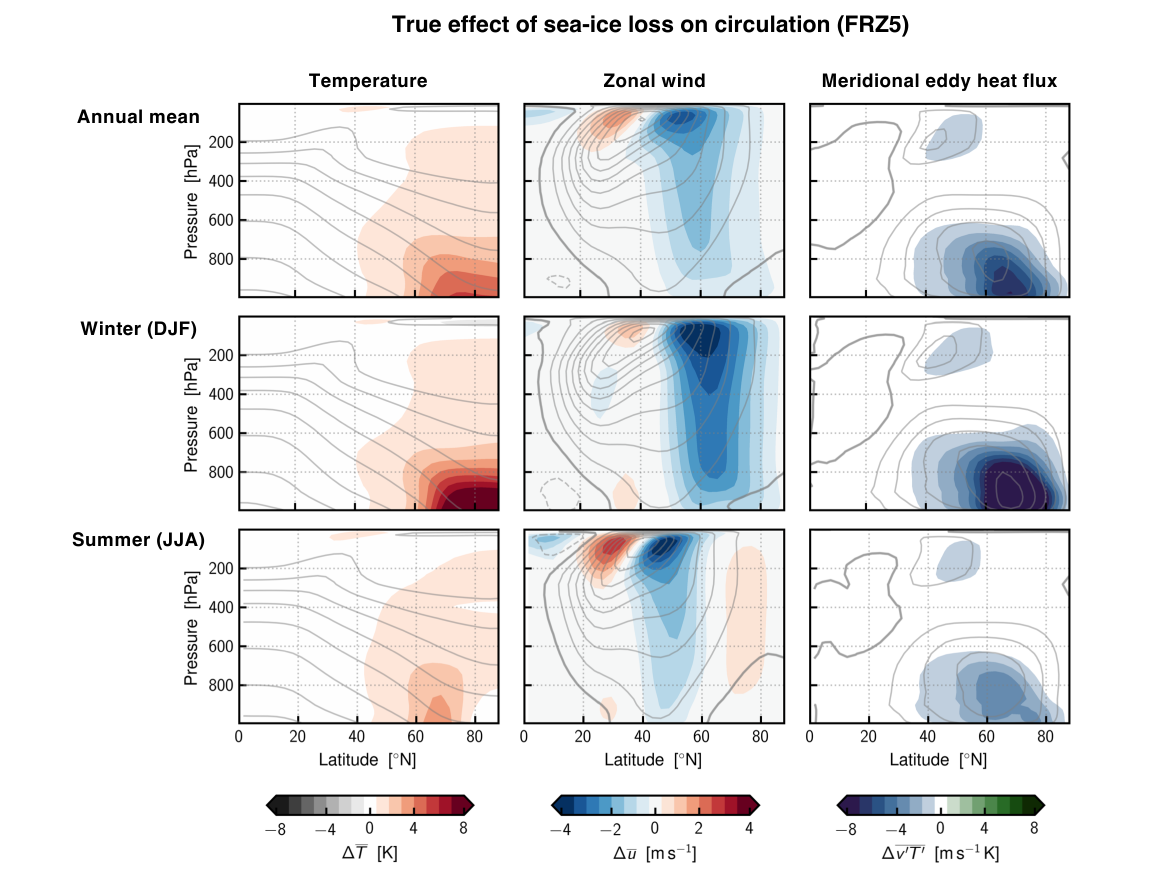}
    \caption{Response of the zonally-averaged circulation to sea-ice loss obtained in the FRZ5 experiment. The top row shows the annual mean response, the middle row shows the response in winter (DJF), and the bottom row shows the response in summer (JJA). The left-hand column shows atmospheric temperature, the central column shows the zonal wind, and the right-hand column shows the meridional eddy heat flux $v^{\prime}T^{\prime}$, which we use here as a measure of storm track intensity. In each panel, the response obtained from taking the difference between FRZ5 and CTRL is shown with colour contours, and the climatology obtained with CTRL is shown with solid contours. Solid contours have increments of $10\,\text{K}$ for temperature (beginning at $300\,\text{K}$), $5\,\text{m\,s}^{-1}$ for zonal wind (with the zero contour marked in bold), and $5\,\text{m\,s}^{-1}\,\text{K}$ for the eddy heat flux (with the zero contour marked in bold).} \label{fig:circ_true_effect}
\end{figure*}

It is interesting to note that between May and July, polar regions in the FRZ$5$ experiment experience cooling relative to the CTRL experiment. This cooling arises because summer sea-ice retreat occurs earlier in the year in the FRZ$5$ experiment, compared with CTRL, and during this period surface temperature increase is temporarily halted as energy input at the surface is used to melt ice instead. This effect can be identified in Figure \ref{fig:true_effect2}, which shows the seasonal cycle of polar surface temperature for the CTRL and FRZ$5$ experiments. The fact that this `latent heating effect' is manifest as a cooling in FRZ$5$, instead of an absence of warming, might appear to be an unphysical artefact of the freezing point modification methodology. However, we do not believe this to be the case, as it is necessary for cooling to occur in FRZ$5$ if the residual warming obtained from ODP$1.1-$FRZ$5$ is to remain polar amplified in summer, which is expected in an idealised, cloud-free GCM \citep{2018JCli...31.5811M,2021GeoRL..4894130F,england2024robust}. Moreover, we note that \citet{chung2023sea} also find sea-ice loss to be associated with cooling in early summer, using an alternative methodology (which does not involve freezing point modification).

Figure \ref{fig:circ_true_effect} shows the effect of sea-ice loss in the FRZ$5$ experiment on the zonally-averaged atmospheric temperature, zonal wind, and meridional eddy heat flux. 
The top row shows the annual mean response, and the middle and bottom rows show the response in DJF and JJA, respectively. The response of atmospheric temperature is polar amplified, and strongest in the lower troposphere (below $p\approx700\,\text{hPa}$).  The response of atmospheric temperature is greatest in winter, and weaker in summer, as was the case for surface temperature. Turning to diagnostics for the atmospheric circulation, we observe that the poleward flank of the eddy driven jet is weakened in response to sea-ice loss, throughout the depth of the troposphere. In the upper troposphere, the jet is additionally strengthened at the edge of the sub-tropics (i.e., the upper-tropospheric jet core is shifted equatorwards). The weakening of the eddy driven jet is accompanied by a weakening of the storm tracks, measured here using the meridional eddy heat flux. As with atmospheric temperature, the zonal wind response in the lower troposphere, and the storm track response, are greatest in winter. These features of the response are generally consistent with results from AGCMs \citep{2022NatCo..13..727S} as well as coupled AOGCMs \citep{2018NatGe..11..155S}. We note that the zonal wind response in the upper troposphere displays less seasonality, but is strongest in summer. 

We note that increasing optical depth in great radiation GCMs tends to induce an equatorward jet shift, even in the absence of sea-ice loss, which is counter to expectations from more sophisticated models \citep{2019JAMES..11..934T,2022JAtS...79..141D}. Due to this deficiency, we do not discuss the circulation response to increasing optical depth as part of our analysis, and instead focus solely on the response to sea-ice loss (spurious or otherwise).

\subsection{Spurious effects of ice-constraining methods on surface temperature}\label{sec:temp}

\begin{figure*}[!t]
    \centering 
    \includegraphics[width=.875\textwidth]{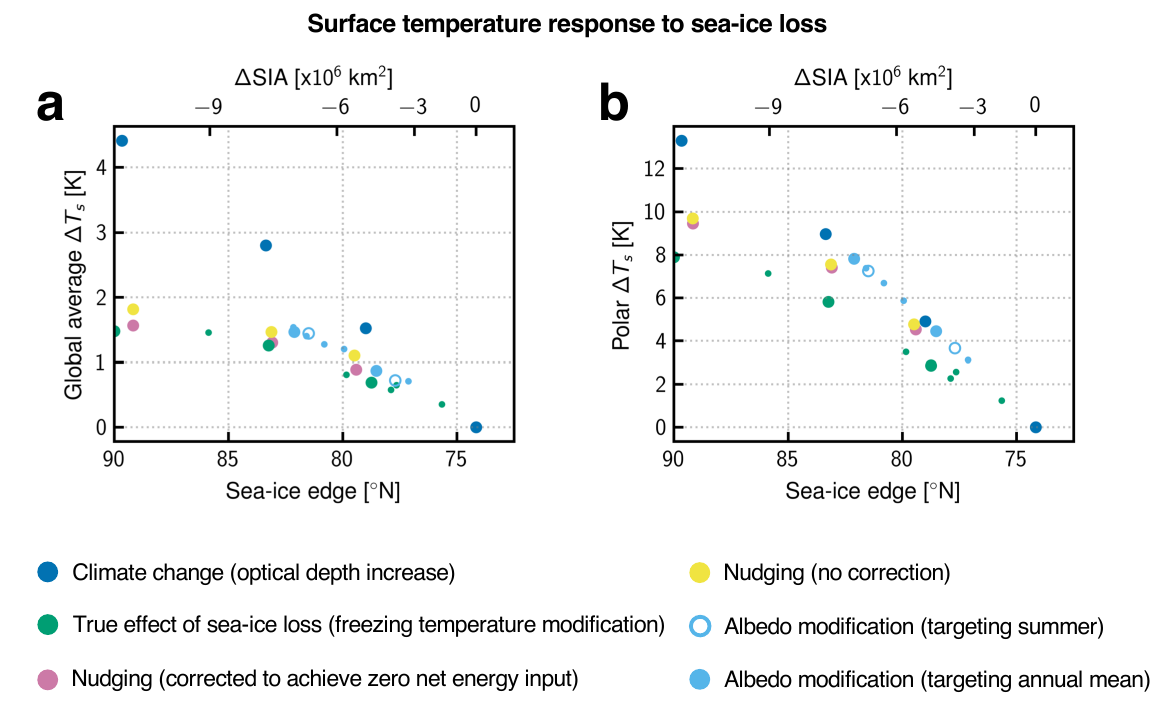}
    \caption{Surface temperature response to sea-ice loss as a function of the sea-ice edge latitude. Panel \textbf{a)} shows annual-mean, globally-averaged surface temperature, and panel \textbf{b)} shows annual-mean, polar-averaged temperature (latitudes $>70^{\circ}\,\text{N}$).  $\Delta$SIA (relative to CTRL) is also shown on the upper $x$-axes. SIA is related to the sea-ice edge latitude via: $\text{SIA}=2\pi a^{2}\left(1-\sin\vartheta_{\text{ice}}\right)$, where $a$ is the Earth's radius. Ice-constrained experiments that best match the SIA obtained in ODP experiments are plotted using a larger marker size. In both panels, the CTRL experiment is shown at $(\Delta\text{SIA}\,,\Delta T)=(0,0)$ (using the same colour as for the ODP experiments). }\label{fig:main_scatter}
\end{figure*}

\subsubsection{Annual mean response}

Area-averaged annual-mean surface temperature responses obtained in each of our experiments, relative to the CTRL experiment, are shown in Figure \ref{fig:main_scatter}. Figure \ref{fig:main_scatter}a shows the response of globally-averaged surface temperature, and Figure \ref{fig:main_scatter}b shows the response of polar-averaged surface temperature (latitudes $>70^{\circ}\,\text{N}$). In each panel, the temperature response is plotted as a function of the sea-ice edge on the lower $x$-axis. This is related to the response of SIA (see caption), denoted $\Delta\text{SIA}$, which is shown on the upper $x$-axis.

\begin{figure*}[!t]
    \centering 
    \includegraphics[width=.975\textwidth]{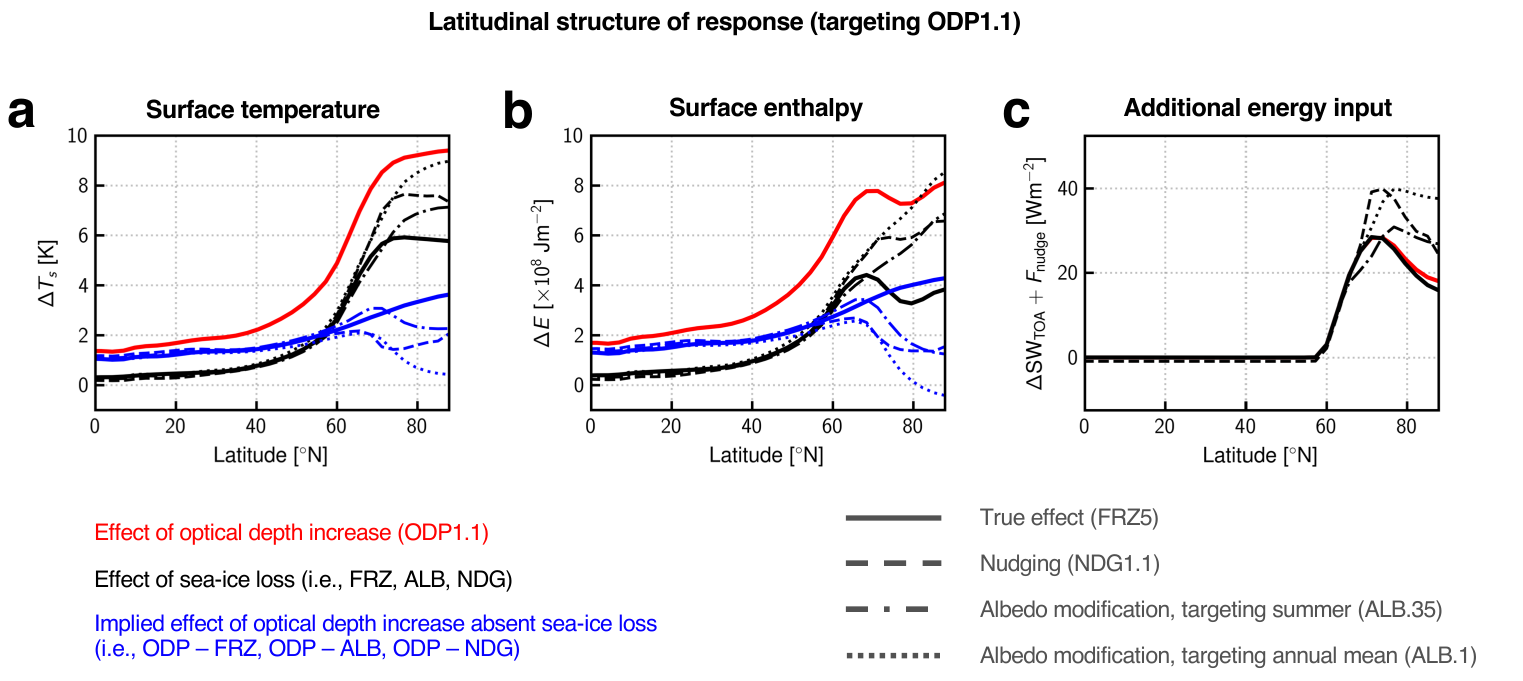}
    \caption{Latitudinal structure of annual-mean response obtained in ODP1.1 and corresponding runs with constrained sea-ice. Panel \textbf{a)} shows the response of surface temperature, panel \textbf{b)} shows the response of surface enthalpy, and panel \textbf{c)} shows the additional energy input for each experiment (relative to CTRL), which is the sum of net top-of-atmosphere (TOA) shortwave response and $F_{\text{nudge}}$ ($F_{\text{nudge}}$ is only non-zero for NDG1.1). The thick red curves show the response to increasing optical depth (ODP-CTRL), thick black curves show the true effect of sea-ice loss obtained from the FRZ5 experiment (FRZ-CTRL), and the thick blue curves show the response of the ODP experiment minus the effect of sea-ice loss (ODP-FRZ). Broken black and blue lines show an equivalent decomposition for the nudging and albedo modification methodologies (using NDG1.1, ALB.35 -- targeting summer, and ALB.1 -- targeting annual mean).} \label{fig:lat_struct}
\end{figure*}

In Figure \ref{fig:main_scatter}a, for a given change in SIA, the smallest surface temperature response to sea-ice loss is obtained by the NDG (pink) and FRZ (true effect of sea-ice loss; green) experiments. In each case, the additional, globally-averaged energy input that causes global mean warming is solely due to the surface albedo response to sea-ice loss (for NDG, this is achieved using the correction term described in Section \ref{sec:expts}\ref{sec:ice_expts}). By contrast, the ALB (light blue) and NDG\_NC (yellow) experiments include an additional, spurious energy input in the global mean, which results in enhanced global warming. This spurious energy input represents the energy required to melt the sea-ice, which itself induces warming (in the corrected NDG experiment, this energy is `borrowed' from the tropics instead).

In Figure \ref{fig:main_scatter}b, the polar surface temperature response in the nudging and albedo modification experiments is comparable to the total surface temperature response due to climate change (obtained in the ODP experiments; dark blue), and significantly greater than that obtained in the freezing point modification experiments, indicating a spurious polar warming effect arising from the nudging and albedo modification methodologies. This effect arises as each of ALB, NDG, and NDG\_NC require surface temperature to be raised above freezing to induce sea-ice loss, before any warming \emph{caused} by sea-ice loss actually occurs. As such, the polar surface temperature response in these experiments represents the combined response to sea-ice loss, plus the energy input required to induce sea-ice loss.

Figure \ref{fig:lat_struct} shows the latitudinal structure of the annual-mean response of surface temperature (panel a) and surface enthalpy (panel b), as well as the additional energy input (relative to CTRL; panel c), for the ODP1.1 experiment, and the constrained sea-ice experiments that target SIA in ODP1.1. 
The surface enthalpy is the prognostic variable at the model's surface, and is defined by \begin{equation}
    E = \begin{cases} -Lh & \text{when}\ h>0, \\ C\left(T_{\text{s}}-T_{\text{freeze}}\right) & \text{when}\ h=0, \end{cases}
\end{equation}
where $h$ is sea-ice thickness, $T_{\text{s}}$ is surface temperature, $T_{\text{freeze}}$ is the freezing point of ice, $L$ is the latent heat of fusion, and $C$ is the heat capacity of the ocean mixed layer (see Section \ref{sec:gcm}). In each panel, the total response to increasing longwave optical depth (ODP$1.1$$-$CTRL) is shown as a solid red line, the true effect of sea-ice loss (FRZ$5$$-$CTRL) is shown as a solid blue line, and the implied effect of increasing optical depth, absent sea-ice loss (ODP$1.1$$-$FRZ$5$), is shown as a solid black line. The responses to sea-ice loss implied by other ice-constraining methodologies are shown as dashed (nudging; NDG$1.1$), dash-dot (albedo targeting summer; ALB$.35$) and dotted (albedo targeting annual mean; ALB$.1$) lines. 

As identified in Section \ref{sec:results}\ref{sec:true_effect}, the surface temperature response in the ODP1.1 experiment is strongly polar amplified and maximal at the pole, as much of the idealised model's climate sensitivity derives from feedbacks associated with sea-ice loss (there are no radiative feedbacks associated with moisture, for example). This effect is amplified by the high-sensitivity of the sea-ice edge to climate change, with each ODP experiment becoming essentially ice-free in summer (high sensitivity of the ice-edge to climate forcings is a feature of aquaplanet GCMs with thermodynamic sea ice, for example CESM2; England \& Feldl \emph{in prep.}). Consequently, much of the polar amplified warming ($\Delta T_{\text{s}}\approx 6\,\text{K}$) is attributed to sea-ice by the FRZ$5$ experiment (solid black line), although the residual warming (solid blue line) is still polar amplified. 
In terms of the surface enthalpy, the response in the ODP1.1 experiment has a secondary peak at the ice-edge (Figure \ref{fig:lat_struct}b, red line). This is co-located with the latitude where the additional energy input due to the effect of sea-ice loss on albedo is maximised (Figure \ref{fig:lat_struct}b and c, solid black lines). This peak is not present in the surface temperature response, demonstrating that some of the additional energy input has gone into melting ice instead of increasing surface temperature.

Each of the experiments that constrain sea-ice while keeping $T_{\text{freeze}}$ unchanged relative to CTRL (NDG1.1, ALB.35, ALB.1) overestimate the surface temperature response to sea-ice loss when compared with FRZ5 (compare the broken black lines with the solid black line in Figure \ref{fig:lat_struct}a). A consequence of this is that the implied warming due to increasing optical depth (broken blue lines) has an unphysical structure that peaks in midlatitudes, before decreasing towards the pole. In the case of the ALB.1 experiment, which targets annually-averaged SIA using albedo modification, this effect is pronounced enough that the residual warming due to increasing optical depth alone is polar de-amplified, which is a clear indicator that too much warming is being attributed to sea-ice loss by this methodology.

Figure \ref{fig:lat_struct}c shows that the nudging and albedo modification experiments include a spurious additional energy input, beyond that which occurs in response to sea-ice loss in either ODP1.1 or FRZ5. This is the spurious energy input required to melt the ice in the absence of either an increase in optical depth, or a modification of the freezing temperature. In the case of the albedo modification experiments, it is largest at the pole, which has the effect of masking the midlatitude local maximum in the surface enthalpy response (compared to the response in FRZ5), and causes the surface temperature response to sea-ice loss to be maximally overestimated at the pole. For the nudging experiment, the spurious additional energy input is more evenly distributed over polar latitudes ($>70^{\circ}\,\text{N}$), which is mirrored by a more even overestimation of the surface enthalpy response in Figure \ref{fig:lat_struct}b. These results are broadly consistent with those obtained by \citet{2022JCli...35.3801E} using a dry EBM.

\subsubsection{Seasonal cycle}

\begin{figure*}[!t]
    \centering 
    \includegraphics[width=.975\textwidth]{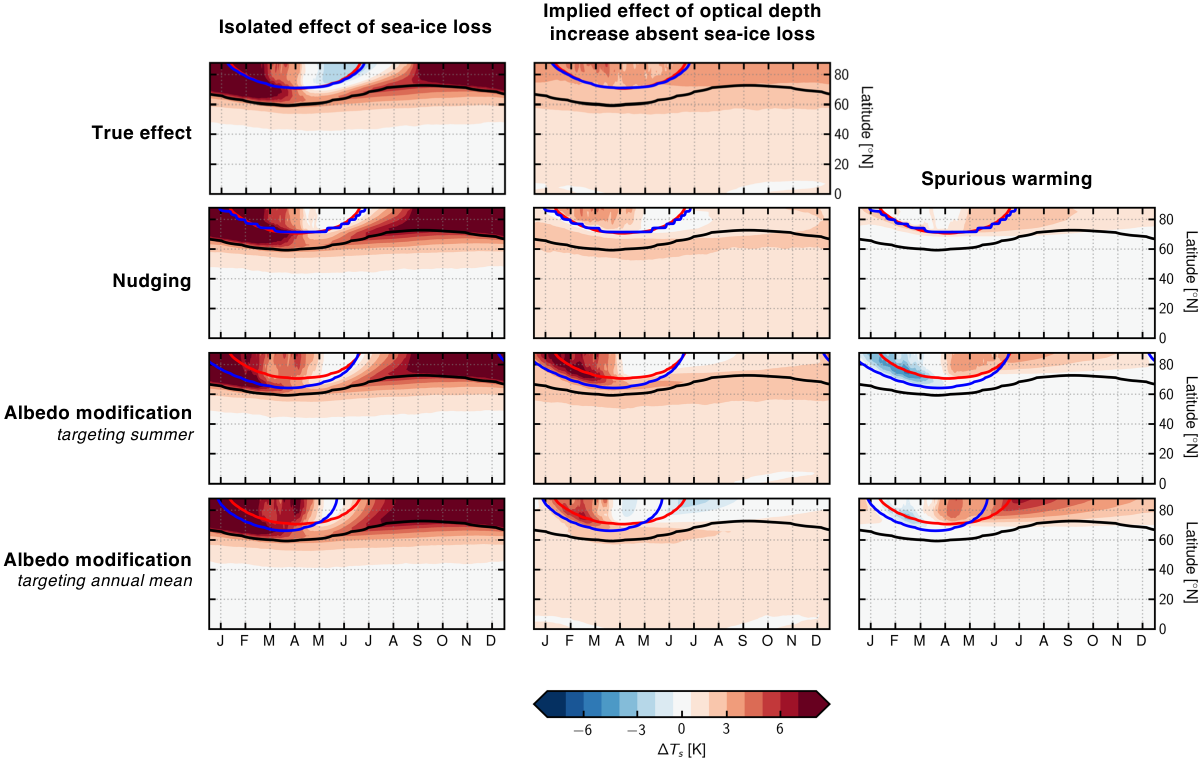}
    \caption{Spurious effects of ice-constraining methods on surface temperature. Left column: Surface temperature response in corresponding experiments with constrained sea-ice loss (FRZ5, NDG1.1, ALB.35 -- targeting summer, and ALB.1 -- targeting annual mean). Centre column: Difference between the total response, and the response obtained in constrained sea-ice experiments (e.g., ODP-NDG). Right column: spurious warming from ice-constraining method, computed as the difference in surface temperature between each experiment targeting ODP1.1 and the experiment FRZ5 (e.g., NDG-FRZ). Red, blue, and black lines represent the sea-ice edge in the ODP1.1, constrained sea-ice (i.e., FRZ, NDG, or ALB, as appropriate), and CTRL experiments, respectively. The data shown in the top row is reproduced from Figure \ref{fig:true_effect}.} \label{fig:multi_contours}
\end{figure*}

\begin{figure}[!t]
    \centering 
    \includegraphics[width=.415\textwidth]{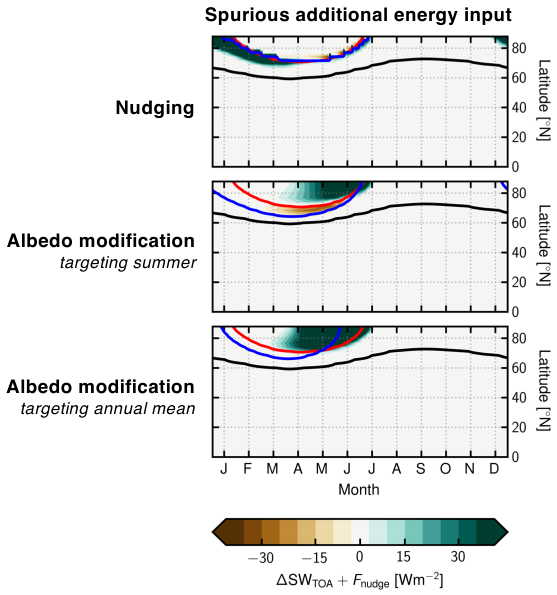}
    \caption{Spurious additional energy input obtained in the NDG1.1, ALB.35 and ALB.1 experiments. This is computed as the sum of the net top-of-atmosphere shortwave response and $F_{\text{nudge}}$, minus the shortwave response in experiment FRZ5 (true effect of sea-ice loss). Red, black, and blue lines show the sea-ice edge, as in Figure \ref{fig:multi_contours}.} \label{fig:spurious_energy}
\end{figure}

To further illustrate the spurious effects of ice-constraining methods on surface temperature, we now examine the seasonal cycle of the temperature response to sea-ice loss obtained in the nudging and albedo modification experiments that target SIA in ODP1.1. This is shown in Figure \ref{fig:multi_contours}. 
At first glance, each method appears to capture a similar warming response to sea-ice loss, compared to the true effect, which is large in autumn, winter and early spring, but suppressed in summer (Figure \ref{fig:multi_contours}, left-hand column). 
However, upon taking the difference between the response obtained in each of NDG1.1, ALB.35, and, ALB.1, and the true response obtained from FRZ5, it can be identified that each of the ice-constraining methods overestimates the temperature response to sea-ice loss during the late summer and early winter (Figure \ref{fig:multi_contours}, right-hand column), while the albedo modification experiments additionally underestimate the response to sea-ice loss during early spring. Consequently, the implied warming in the absence of sea-ice loss (Figure \ref{fig:multi_contours}, central column) exhibits an unphysical seasonal dependence which cannot be explained by any of the physical processes included in the idealised GCM. This unphysical seasonality is not present when the true effect of sea-ice loss is isolated using freezing point modification (Figure \ref{fig:true_effect}, right-panel), or consistent with the results of previous work \citep{chung2023sea}.

The spurious additional energy input in each of these experiments, 
associated with the nudging and albedo modification methodologies, is shown in Figure \ref{fig:spurious_energy}. For the nudging experiment, the additional term in Equation \ref{eq:nudge} causes additional energy input during winter and early spring, in order to prevent sea-ice from growing beyond the target SIA. This energy initially goes into warming the mixed layer, which subsequently becomes ice covered, and it is only when sea-ice retreat occurs in summer that the additional energy input manifests as spurious increase in surface temperature. By contrast, albedo modification has no effect on the top-of-atmosphere energy balance during the polar night, which means that spurious additional energy input occurs later in the year (relative to the nudging experiment), in late spring and early summer. However, because the spurious energy input occurs when the surface is already ice covered, the temperature response to this forcing emerges at the surface immediately, and is then communicated to the mixed layer via the conductive heat flux through the ice. Consequently, both nudging and albedo modification cause spurious warming that is greatest in summer, despite the fact that spurious energy input occurs at different times of the year.

As noted above, the albedo modification experiments additionally underestimate the surface temperature response during late winter. This effect arises because these experiments retain too much sea-ice during winter, relative to the target climate change experiment. When sea-ice is present, the effective heat capacity of the surface is reduced \citep{2022JCli...35.1629H}, which causes the surface temperature to cool more in winter than would be the case if it were ice-free.

\begin{figure*}[!t]
    \centering 
    \includegraphics[width=.855\textwidth]{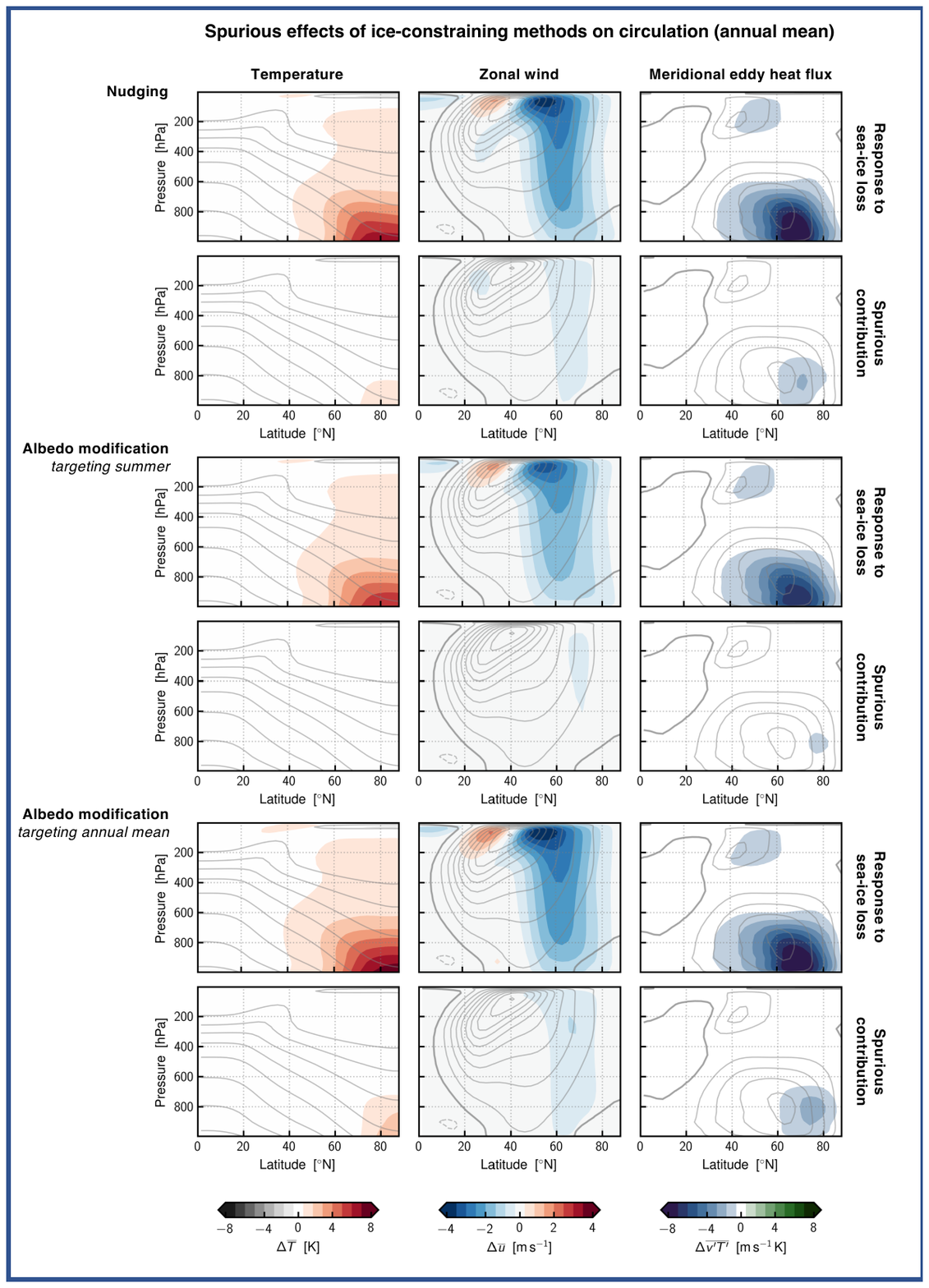}
    \caption{Annually-averaged effect of sea-ice loss on circulation for experiments targeting SIA in the ODP1.1 experiment. For each variable, the response to sea-ice loss is computed relative to the CTRL experiment, and the spurious response is computed relative to the FRZ5 experiment. Solid contours show the climatology obtained with the CTRL experiment (see Figure \ref{fig:circ_true_effect} caption for more information).} \label{fig:spurious_circ}
\end{figure*}

\subsection{Impact on large-scale atmospheric circulation}

In this section, we analyse the spurious impacts of the nudging and albedo modification methodologies on the circulation response to sea-ice loss. 

\subsubsection{Annual- and zonal-mean circulation response}

Figure \ref{fig:spurious_circ} shows the annual mean response of the zonally-averaged circulation to sea-ice loss for the nudging and albedo modification experiments that target SIA obtained in the ODP1.1 experiment, relative to CTRL, as well as the difference between each response and the true effect of sea-ice loss (e.g., NDG1.1$-$FRZ5). The full response is shown in the upper row for each experiment, and the spurious contribution is shown in the lower row for each experiment. The left-hand column shows atmospheric temperature, the central column shows the zonal wind, and the right-hand column shows the meridional eddy heat flux, $\overline{v^{\prime}T^{\prime}}$ (the overline denotes a zonal and day-of-year average, and primes indicate departures therefrom), which we use as a simple measure of storm track intensity. 

For each experiment, the spurious temperature response in Figure \ref{fig:spurious_circ} has a magnitude similar to that obtained for surface temperature (see, e.g., Figure \ref{fig:lat_struct}a, comparing the broken and solid black lines). It is mostly confined to the lower troposphere ($p\gtrsim700\,\text{hPa}$), and at greater altitude the spurious temperature response is weak compared to the total response. It is notable that the `mini global warming' response to sea-ice loss obtained in AOGCMs with constrained sea-ice \citep{2015JCli...28.2168D} cannot be identified in Figure \ref{fig:spurious_circ}. In AOGCMs, the influence of sea-ice loss on the tropics is driven by the response of ocean dynamics \citep{2016JCli...29.6841T,2018GeoRL..45.4264W,2020NatGe..13..275E}, which cannot occur in the idealised GCM as it is configured using a slab ocean with prescribed ocean heat transport.

Due to spurious warming, the nudging and albedo modification experiments overestimate the response of the zonal-mean zonal wind to sea-ice loss. This effect is relatively pronounced for the NDG1.1 and ALB.1 experiments, where there is a spurious additional weakening of the zonal wind around $65^{\circ}\,\text{N}$, which is roughly barotropic, and near the surface accounts for approximately $35\%$ of the total zonal wind response to sea-ice loss. In addition to the zonal wind, nudging and albedo modification also drive a spurious additional weakening in storm track activity, measured using the meridional eddy heat flux. As with the temperature response, this effect is mostly confined to the lower troposphere, and is relatively weak compared with the total response to sea-ice loss suggested by each method. 

\subsubsection{Storm track response}

\begin{figure}[!t]
    \centering 
    \includegraphics[width=.475\textwidth]{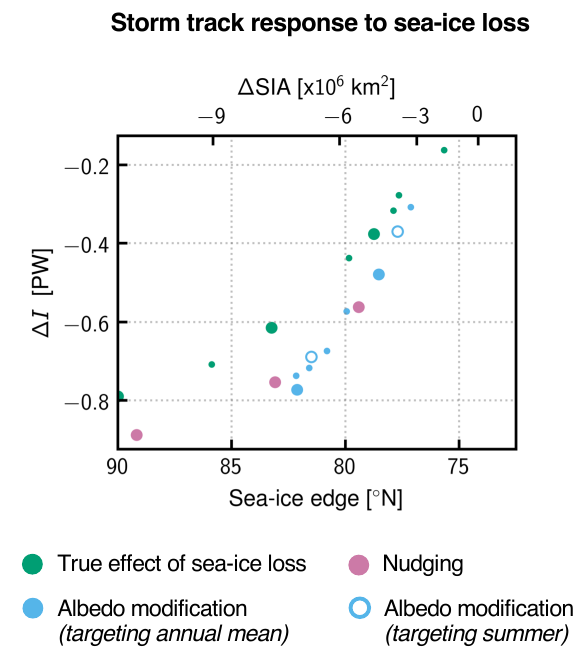}
    \caption{Storm track response to sea-ice loss, quantified using the storm track intensity $I$ (Equation \ref{eq:I}). Ice-constrained experiments that best match the target SIA obtained in the ODP1.05, ODP1.1, and ODP1.2 experiments are plotted using a larger marker size.} \label{fig:storm_track}
\end{figure}

To analyse the spurious effects of nudging and albedo modification on the storm track in greater detail, we consider an additional metric for storm track activity, namely the `storm track intensity' $I$ introduced by \citet{2018JAtS...75.1979S}. Storm track intensity is defined by \begin{equation}
    I(\vartheta)=2\pi a\cos\vartheta F_{\text{TE}}(\vartheta), \label{eq:I}
\end{equation}
where $F_{\text{TE}}=\overline{v^{\prime}m^{\prime}}$ is the meridional eddy moist static energy (MSE) flux. The MSE itself is given by $m=c_p T+gz+Lq$, where $T$ is temperature, $z$ is geopotential height, and $q$ is specific humidity, $c_p$ is the specific heat capacity of dry air, and $g$ is the acceleration due to gravity.

Figure \ref{fig:storm_track} shows the response of storm track intensity, $\Delta I$, to sea-ice loss for the FRZ, NDG, and ALB experiments, relative to CTRL, plotted against the latitude of the sea-ice edge. In this figure, $\Delta I$ is integrated vertically (mass weighted) over the depth of the atmosphere, and meridionally-averaged between $30^{\circ}\,\text{N}$ and $60^{\circ}\,\text{N}$. As with the eddy heat flux shown in Figure \ref{fig:spurious_circ}, sea-ice loss drives a decrease in storm track intensity, which is enhanced in the nudging and albedo modification experiments compared with the true response to sea-ice loss due to spurious energy input.

Using the MSE budget framework described by \citet{2018JAtS...75.1979S} and \citet{2022JCli...35.2633S}, it is possible to derive an equation for the spurious response of the storm track intensity, $\Delta I^{\ast}$, to spurious energy input, $\Delta I^{\ast}_{\text{input}}$, associated with the nudging and albedo modification methodologies (see Appendix for derivation; asterisks indicate a spurious forcing or response): \begin{equation}
    \Delta I^{\ast} = \Delta I^{\ast}_{\text{LW}} - \Delta I^{\ast}_{\text{M}} + \Delta I^{\ast}_{\text{input}}. \label{eq:I_eq2}
\end{equation}
Above, $\Delta I^{\ast}_{\text{LW}}=2\pi a^{2}\int^{\vartheta}_{\nicefrac{\pi}{2}}\Delta \text{LW}_{\text{TOA}}\,\cos\vartheta\text{d}\vartheta$ is the spurious longwave cooling response, where $\text{LW}_{\text{TOA}}$ is the net downward top-of-atmosphere longwave radiative flux, $\Delta I^{\ast}_{\text{M}}=2\pi a \cos\vartheta \Delta F_{\text{M}}(\vartheta)$ is the spurious response of heat transport by the mean flow, where $F_{\text{M}}=\overline{v}\,\overline{m}$. To compute the spurious responses $\Delta I^{\ast}_{\text{LW}}$ and $I^{\ast}_{\text{M}}$, the differences $\Delta \text{LW}_{\text{TOA}}$ and $\Delta F_{\text{M}}$ are taken between sea-ice loss experiments with additional heat (i.e., NDG and ALB) and the FRZ experiment that captures the true response (absent additional heat). The spurious forcing $\Delta I^{\ast}_{\text{input}}$ is given by: \begin{equation}
\Delta I^{\ast}_{\text{input}} = 2\pi a^{2} \int^{\vartheta}_{\frac{\pi}{2}} \left(F_{\text{nudge}} -\Delta\alpha_{\text{sfc}} F^{\downarrow}_{\text{sw, sfc}}\right)\cos\vartheta\text{d}\theta, \label{eq:Ispur}
\end{equation}
where non-zero $\Delta\alpha_{\text{sfc}}$ between the ALB and FRZ experiments is entirely spurious (e.g., due to artifical darkening of the ice). Equation \ref{eq:I_eq2} predicts that changes in storm track intensity are due to changes in the meridional flux divergence implied by spurious energy input, integrated over latitude \citep{2022JCli...35.2633S}. 

\begin{figure}[!t]
    \centering 
    \includegraphics[width=.475\textwidth]{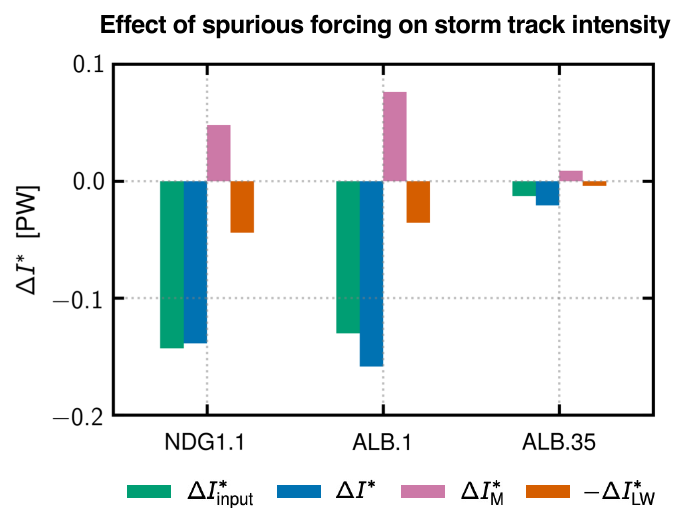}
    \caption{Response of different terms in the MSE intensity budget to spurious forcing in the NDG1.1, ALB.1, and ALB.35 experiments. The green bar shows the spurious forcing, the blue bar shows the response of eddy MSE intensity, the pink bar shows the response of the mean meridional circulation MSE intensity, and orange bar shows the response of radiative cooling. All quantities are meridionally averaged between $30^{\circ}\text{N}$ and $60^{\circ}\text{N}$. } \label{fig:spurious_track}
\end{figure}

Figure \ref{fig:spurious_track} shows the spurious forcing $\Delta I^{\ast}_{\text{input}}$ (green), and the spurious responses $\Delta I^{\ast}$ (blue), $\Delta I^{\ast}_{\text{M}}$ (pink), and $-\Delta I^{\ast}_{\text{LW}}$ (orange), obtained by taking the difference between the nudging and albedo modification experiments targeting SIA in ODP1.1, and the true effect of sea-ice loss in ODP1.1 (given by FRZ5). In each case, the MSE framework shows that spurious forcing, $\Delta I^{\ast}_{\text{input}}$, drives a spurious storm track response, $\Delta I^{\ast}$, of a similar magnitude. The storm track response is partially offset by an increase in radiative cooling, $\Delta I^{\ast}_{\text{LW}}$ in response to spurious forcing, but amplified by a weakening of the midlatitude mean meridional circulation (Ferrel cell), which transports heat equatorwards in midlatitudes. As the Ferrell cell is eddy-driven \citep{2017aofd.book.....V}, this effect could be interpreted as an eddy-feedback, whereby the initial weakening of transient eddies in response to spurious energy input causes a weakening of the mean meridional circulation, which subsequently drives a further response in the transient eddies. 

\begin{figure*}[!t]
    \centering 
    \includegraphics[width=.855\textwidth]{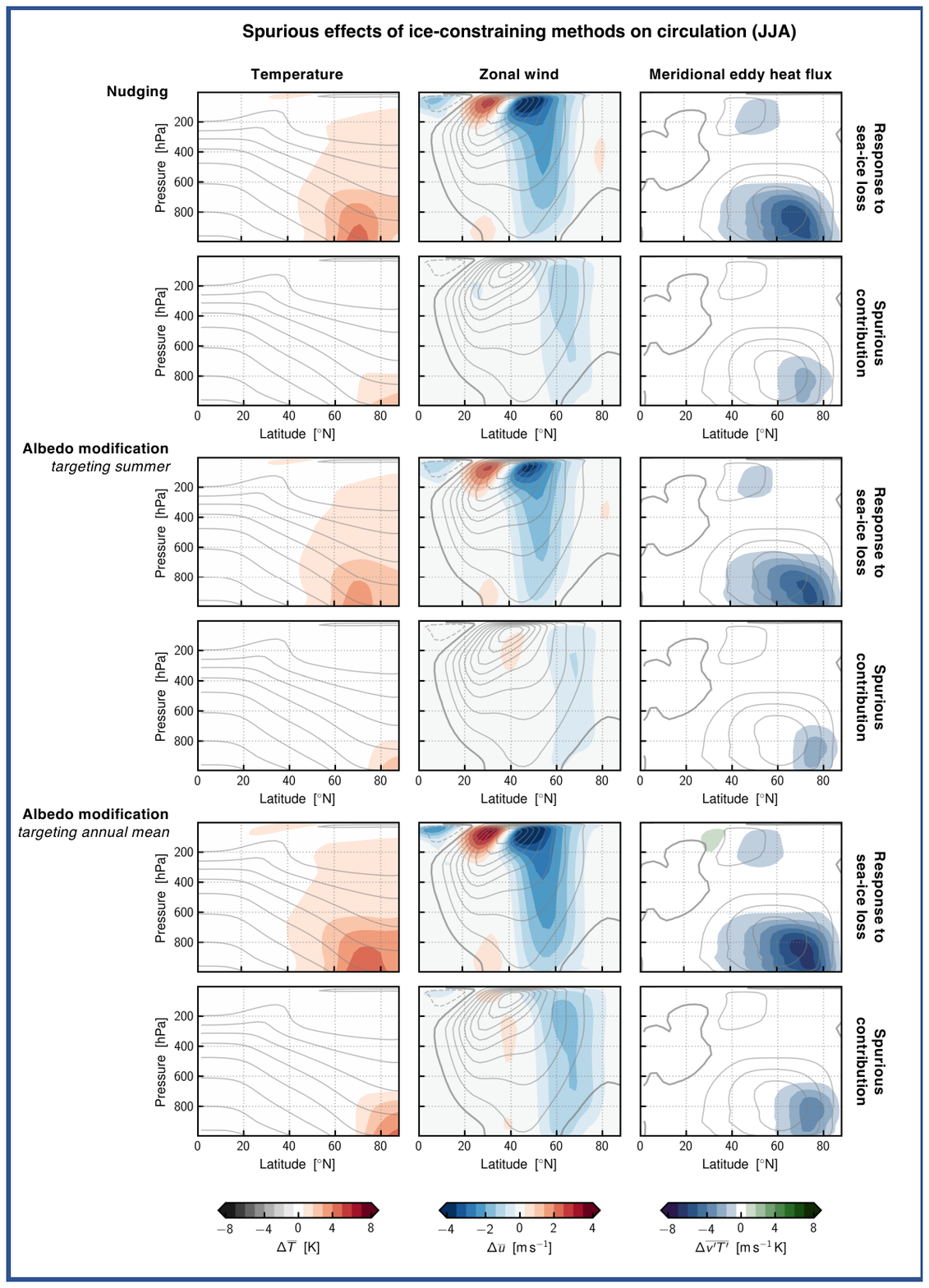}
    \caption{Effect of sea-ice loss atmospheric circulation in summer (JJA), for experiments targeting SIA in the ODP1.1 experiment. For each variable, the response to sea-ice loss is computed relative to the CTRL experiment, and the spurious response is computed relative to the FRZ5 experiment. Solid contours show the JJA climatology obtained with the CTRL experiment (see Figure \ref{fig:circ_true_effect} caption for more information).} \label{fig:spurious_circ_jja}
\end{figure*}

\subsubsection{Zonal-mean circulation response in JJA}

Finally, we consider how the spurious effects of ice-constraining methods on circulation vary with the seasonal cycle. Figure \ref{fig:spurious_circ_jja} shows the same information as Figure \ref{fig:spurious_circ}, but now averaged over boreal summer (JJA) as opposed to the whole year. In summer, the true response of atmospheric circulation to sea-ice loss was found to be weaker than the response in either the annual mean or winter (see Figure \ref{fig:circ_true_effect}). However, this is less obvious when inspecting the total response to sea-ice loss obtained using the nudging or albedo modification methods, presented in Figure \ref{fig:spurious_circ_jja}. This is because the spurious circulation response is larger in summer compared with the annual mean (consistent with the surface temperature response discussed in Section \ref{sec:temp}), which causes the total JJA circulation response to sea-ice loss in these experiments to be substantially overestimated. 

As with surface temperature (see Figure \ref{fig:multi_contours}), the spurious atmospheric temperature responses in Figure \ref{fig:spurious_circ_jja} are largest at the pole, and offset from the true temperature response (Figure \ref{fig:circ_true_effect}) which is greatest at around $70^{\circ}\,\text{N}$ in summer. This drives a spurious weakening of the zonal-mean zonal wind around $70^{\circ}\,\text{N}$, masking the true effect of sea-ice loss, which actually strengthens the zonal-wind at high-latitudes in FRZ5. As with the zonal wind response, the nudging and albedo modification experiments overestimate the meridional eddy heat flux response, with a spurious contribution that is larger in summer compared with the annual mean. At the surface, this effect accounts for roughly $45\%$ of the total response to sea-ice loss obtained in the NDG1.1 and ALB.1 experiments.

\section{Conclusions and Discussion}\label{sec:discuss}

In this study, we have analysed the effect of sea-ice loss on the climate of an idealised GCM with thermodynamic sea-ice. We use three methods to constrain sea-ice: freezing point modification, which isolates the true response to sea-ice loss in our model, i.e., the response obtained without using artificial energy input to melt ice, and nudging and albedo modification, which are commonly used methods for constraining sea-ice in fully coupled AOGCMs (see, e.g., \citealp{2020GeoRL..4785788S}). 

We show that nudging and albedo modification cause too much warming in response to sea-ice loss, when compared with the true effect isolated using freezing point modification (e.g., Figure \ref{fig:main_scatter}a). This arises because the surface temperature response to sea-ice loss induced using these methods includes a spurious contribution, that is a direct effect of the energy input required to melt sea-ice, rather than an effect of sea-ice loss itself. We find that spurious warming is most prominent in polar regions, where, in the most extreme cases, it causes the temperature response to sea-ice loss to be comparable to the temperature response to climate change (Figure \ref{fig:main_scatter}b, Figure \ref{fig:lat_struct}a), which would attribute no polar warming to other factors (e.g., GHG emissions).
This arises as the spurious additional energy input associated with nudging, and to a greater extent albedo modification, increases with latitude towards the pole (Figure \ref{fig:lat_struct}c). These results confirm and extend those presented by \citet{2022JCli...35.3801E} and \citet{2024ERCli...3a5003F}. Finally, we show that spurious warming associated with both nudging and albedo modification is maximal in boreal summer, in spite of the fact that the true temperature response to sea-ice loss is greatest in boreal winter (Figure \ref{fig:true_effect}b).

Using freezing point modification, we show that the true effect of sea-ice loss on circulation in our idealised GCM primarily consists of a weakening of the zonal-mean zonal wind on the poleward flank of the midlatitude jet, accompanied by a strengthening  on the equatorward flank of the jet in the upper troposphere, and a weakening of the midlatitude storm track (Figure \ref{fig:circ_true_effect}). This response is greatest in winter, when the effect of sea-ice loss on surface temperature is greatest. Spurious warming in nudging and albedo modification experiments leads to an overestimation of the circulation response to sea-ice loss compared to the true effect (Figure \ref{fig:spurious_circ}). Using the MSE budget framework proposed by \citet{2018JAtS...75.1979S} and \citet{2022JCli...35.2633S}, we show that spurious weakening of storm track intensity in nudging and albedo modification experiments occurs in response to a reduction in the divergence of radiative and non-atmospheric energy fluxes, implied by spurious energy input at high latitudes (Figure \ref{fig:storm_track}). As with the temperature response, we find that the effect of spurious energy input on the circulation is greatest in boreal summer (Figure \ref{fig:spurious_circ_jja}). In some sense, this is a desirable result, given that most interest is in the wintertime, when the response to sea-ice loss is believed to be largest \citep{2010JCli...23..333D}.

Our results suggest that coupled AOGCMs that use nudging or albedo modification to constrain Arctic sea-ice may overestimate the response of temperature and circulation, particularly during boreal summer where the circulation response contains a large spurious contribution (Figure \ref{fig:spurious_circ_jja}).   However, it is important to note that our results will be sensitive to the idealised model configuration we use. In particular, the omission of key climate processes, such as the cloud and water vapour radiative feedbacks, means that the ice-albedo feedback has an outsized influence on the climate of the idealised model. Further, we note that our model does not exhibit a prominent `mini global warming response', identified in studies that use coupled AOGCMs \citep{2015JCli...28.2168D}, and specifically, there is little warming in any of our experiments in the tropical free troposphere. This likely arises because our model does not include a dynamic ocean \citep{2016JCli...29.6841T,2018GeoRL..45.4264W,2020NatGe..13..275E}.  Therefore, we cannot comment on the extent to which this tropical response is a `real' effect of coupling, or a spurious effect of the methods used to constrain sea-ice in AOGCMs.

\begin{figure*}[!t]
    \centering 
    \includegraphics[width=.82\textwidth]{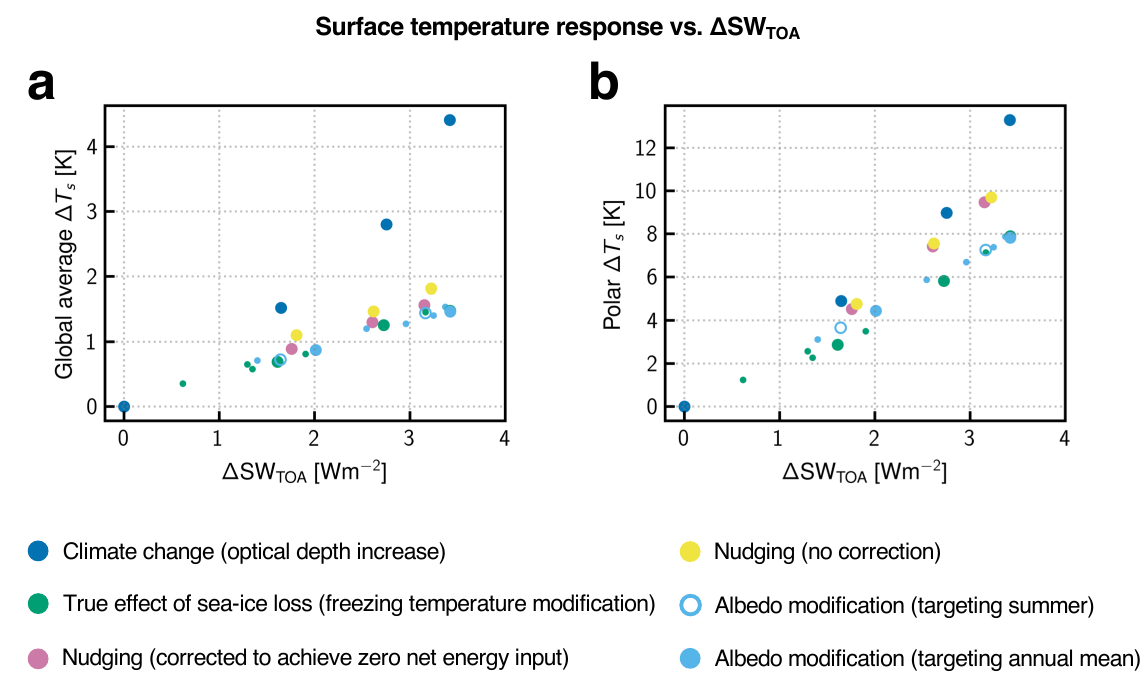}
    \caption{Surface temperature response to sea-ice loss as a function of the change in globally-averaged net top-of-atmosphere (TOA) shortwave radiation. Panel \textbf{a)} shows annual-mean, globally-averaged surface temperature, and panel \textbf{b)} shows annual-mean, polar-averaged temperature (latitudes $>70^{\circ}\,\text{N}$).  As in Figure \ref{fig:main_scatter}, ice-constrained experiments that best match the SIA obtained in ODP experiments are plotted using a larger marker size. }\label{fig:main_scatter2}
\end{figure*}

Given these limitations, separating the `true effect' of sea-ice loss from the spurious effects of nudging and albedo modification in sophisticated AOGCMs remains an important topic for future research. Unfortunately, freezing point modification cannot be trivially implemented in AOGCMs, as it would require modification of the equation of state used in either the ocean or sea-ice component of the model. {
Nevertheless, it is worth noting that our albedo and freezing point modification experiments yield roughly the same globally-averaged, and polar-averaged, surface temperature responses as a function of the change in top-of-atmosphere shortwave, $\Delta \text{SW}_{\text{TOA}}$, which is shown in Figure \ref{fig:main_scatter2}. A similar result is obtained for the storm track intensity, shown in Figure \ref{fig:main_scatter3}. This suggests an alternate strategy for albedo modification, namely targeting the total effect of sea-ice loss on albedo, instead of sea-ice area or volume itself, which in our simplified model would yield the true temperature response to sea-ice loss. 

As discussed in the introduction, \citet{2024ERCli...3a5003F} suggest that multi-parameter pattern scaling can be utilised to correct for spurious heating in AOGCM simulations that use albedo modification to constrain sea-ice. This is achieved by defining the climate response to sea-ice loss so that it scales with $\text{SW}_{\text{TOA}}$ instead of sea-ice area. The fact that the response to sea-ice loss in our FRZ (true effect) and ALB experiments collapse onto the same curve, when plotted as a function of $\Delta\text{SW}_{\text{TOA}}$, supports this choice. 

\begin{figure}[!t]
    \centering 
    \includegraphics[width=.42\textwidth]{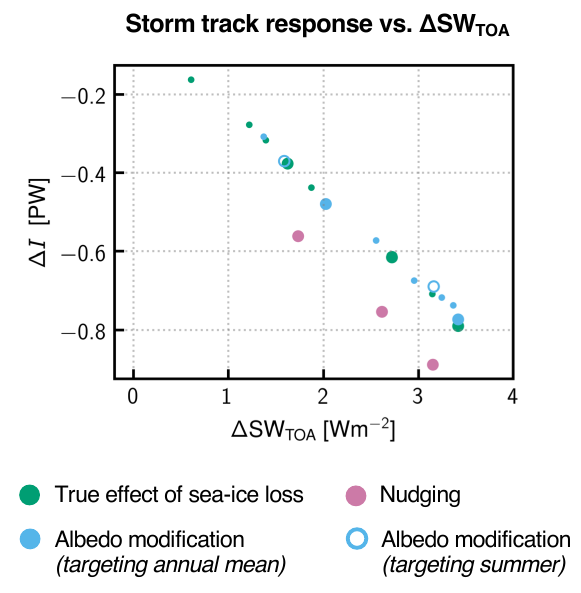}
    \caption{Response of storm track intensity, $I$, to sea-ice loss, as a function of the change in globally-averaged net top-of-atmosphere (TOA) shortwave radiation.   As in Figure \ref{fig:spurious_track}, ice-constrained experiments that best match the target SIA obtained in the ODP1.05, ODP1.1, and ODP1.2 experiments are plotted using a larger marker size.}\label{fig:main_scatter3}
\end{figure}

However, this methodology also has some limitations. First, \citeauthor{2024ERCli...3a5003F} show that their method is not exact; specifically, by applying their methodology to the EBM used by \citet{2022JCli...35.3801E}, they find that the surface temperature response corrected for artifical heat approaches the known true response to sea-ice loss in the EBM, but does not equal it. The magnitude of spurious warming is underestimated at high-latitudes, and notably this issue appears to become worse when moisture is included in the model ($\approx 0.5\,\text{K}$ underestimation in the dry EBM compared with $\approx 1\,\text{K}$ in the moist EBM). Additionally, the seasonal cycle of spurious warming isolated using this methodology is greatest in winter, rather than summer (Luke Fraser--Leach, \emph{personal communication}), which is inconsistent with the results obtained using our idealised model, as well as the fact that albedo modification causes spurious energy input in summer (which should immediately warm the ice-surface temperature). While beyond the scope of this study, we believe it would be useful to test the pattern scaling approach using the idealised model experiments we have presented. Finally, \citeauthor{2024ERCli...3a5003F} note that $F_{\text{TOA}}$ is a less appropriate choice for the scaling variable when correcting for artificial heat in AOGCM simulations that use methods other than albedo modification, and that the `correct' choice for an alternative scaling variable is not immediately obvious. This problem is evident in Figures \ref{fig:main_scatter2}b and \ref{fig:main_scatter3}, where the nudging experiments exhibit too great a response as a function of $\Delta \text{SW}_{\text{TOA}}$. This is because $\Delta \text{SW}_{\text{TOA}}$ does not account for the spurious energy that is input by the nudging methodology at high latitudes.}

{
In this paper, we have focused on the spurious impacts of ice-constraining methods on the response of coupled AOGCMs to sea-ice loss. However, it is worth considering the implications of our results, as well as those of \citet{2022JCli...35.3801E} and \citet{2024ERCli...3a5003F}, for studies that make use of AGCMs with prescribed SST and SIC. Whether AGCM experiments contain a spurious contribution will depend on whether the SST prescribed in regions of sea-ice loss is attributable to sea-ice loss itself, or if it is actually the cause of sea-ice loss. Typically, studies set newly ice-free SST to the freezing point of salt water (e.g., \citealp{2004JCli...17..857M,2006JCli...19.1109S,2010JCli...23..333D,2015JGRD..120.3209N}), or prescribe SSTs obtained from simulations of climate change (e.g., \citealp{2013JCli...26.1230S,2014JCli...27..244P,2014NatCo...5.4646K,2015JCli...28.7824S,2019GMD....12.1139S}). If applied to our idealised model, each of these approaches would prescribe warmer SSTs than the newly ice-free SST obtained in our FRZ experiments, which is sub-freezing at the ice-edge by definition (because the freezing point has been reduced to induce sea-ice loss). This implies that AGCM experiments may also include a spurious warming in polar regions, which will be exaggerated in cases where future SST is prescribed when sea-ice is lost (notably including PAMIP; \citealp{2019GMD....12.1139S}). 

Our results suggest that climate model simulations with constrained sea-ice should be treated with caution. However, it is important to emphasise that across all of our idealised simulations, both with and without a spurious effect, the characteristics of the zonally-averaged temperature and atmospheric circulation response to sea-ice loss are largely the same, especially in DJF when the true response is greatest compared with other times of the year (Figures \ref{fig:circ_true_effect} and \ref{fig:spurious_circ}). In each case, temperature increases at high-latitudes in the lower troposphere, and the eddy driven jet weakens on its poleward flank, accompanied by a reduction in storm track activity. These results are broadly consistent with those obtained using sophisticated AGCMs (e.g., \citealp{2022NatCo..13..727S}) and AOGCMs (e.g., \citealp{2018NatGe..11..155S}). Spurious warming associated with ice-constraining methods serves mostly to increase the magnitude of the response, and it is not the only cause of uncertainty in this respect. For example, there is significant spread in the magnitude of the mid-latitude jet response obtained with AGCMs, potentially due to variation in the strength of eddy-feedbacks between models \citep{2022NatCo..13..727S}. In addition, \citet{2014JCli...27..244P} highlight that different representations of clouds and aerosols in models may have a significant impact on the magnitude of the stratospheric polar vortex response to sea-ice loss. Other factors that contribute to model spread include differences in the background state \citep{2017JCli...30.4547S} and internal variability, especially in the stratosphere \citep{2021JCli...34.3751P,2022JCli...35.3109S}. In order to more precisely constrain the circulation response to Arctic sea-ice loss, it is equally important that sources of intermodel variability, and the effects of spurious heating in experiments with constrained or prescribed sea-ice, are better understood.}

Finally, it is important to note that the term `spurious' is open to interpretation. \citet{2013JCli...26.1230S} argue that the emergence of newly ice-free areas with warmed SSTs is inseparable from sea-ice loss, and so should be included in simulations which investigate the climate response to sea-ice loss. Within this context, it may be argued that the appropriateness (or otherwise) of current AGCM and AOGCM experiments depends on the framing of the science question under consideration. Specifically, existing experiments with constrained sea-ice implicitly target the following question: \begin{quote}
     {Q1. What is the difference between the current climate and a counterfactual climate where sea-ice is artificially melted?}
\end{quote}
Our intention is not to suggest that studying this question is without merit, but instead to stress that \emph{it is important to acknowledge the nuance that differentiates this question from the alternate question}: \begin{quote}
    {Q2. What is the effect of sea-ice loss on climate?}
\end{quote} which many, existing studies claim to address.

\acknowledgments
We thank Paul Kushner and Luke Fraser--Leach for engaging in discussion that benefited this work. NTL, JAS, RG, WJMS, and SIT acknowledge support from the Natural Environment Research Council (NERC) under grant agreement ArctiCONNECT (NE/V005855/1). MRE is supported by an 1851 Research Fellowship.  RM is funded by a NERC GW4+ Doctoral Training Partnership studentship (NE/S007504/1).

%
%

%
\appendix

\appendixtitle{Storm track response to spurious energy input}

To interpret the effect of spurious energy input on the storm track intensity response $\Delta I$, we make use of the MSE framework described by \citet{2018JAtS...75.1979S} and \citet{2022JCli...35.2633S}. The meridional MSE budget is given by: \begin{equation}
\nabla\cdot F_{\text{TE}} = \text{Ra} + \text{TF} - \nabla\cdot F_{\text{M}}, \label{eq:mse_budget}
\end{equation}
where $\text{Ra}$ is radiative cooling (the difference between net downward top of atmosphere and surface radiative fluxes), $\text{TF}$ is the surface turbulent flux into the atmosphere, and $F_{\text{M}}=\overline{v}\,\overline{m}$ is the meridional MSE flux due to the mean circulation. Above, $\text{Ra}$ and $\text{TF}$ have their global-mean removed, following \citet{2008JCli...21.3521K} and \citet{2022JCli...35.2633S}. By making use of the surface energy budget \begin{equation}
\text{TF} = \text{SW}_{\text{s}} + \text{LW}_{\text{s}} - \nabla\cdot F_{\text{NA}}, 
\end{equation}
Equation \ref{eq:mse_budget} can be re-written as \begin{equation}
\nabla\cdot F_{\text{TE}} = \text{SW}_{\text{TOA}} + \text{LW}_{\text{TOA}} - \nabla\cdot F_{\text{M}} - \nabla\cdot F_{\text{NA}}, \label{eq:mse_budget2}
\end{equation} 
where $\text{SW}_{\text{TOA}}$ and  $\text{LW}_{\text{TOA}}$ are the net downward top of atmosphere shortwave and longwave radiative fluxes, and $\nabla\cdot F_{\text{NA}}$ denotes energy flux divergence due to non-atmospheric processes, such as ocean heat transport and, in the case of the NDG experiments, artificial energy input used to remove sea-ice.

Using Equation \ref{eq:mse_budget2}, a change in storm track intensity between climates can be written as \begin{equation}
\Delta I = \Delta I_{\text{SW}} + \Delta I_{\text{LW}} - \Delta I_{\text{M}} - \Delta I_{\text{NA}}, \label{eq:I_eq}
\end{equation}
by meridionally integrating Equation \ref{eq:mse_budget2} from the pole to a latitude $\vartheta$ ($\Delta I_{\text{SW}}$ and $\Delta I_{\text{LW}}$ have their global mean removed; \citealp{2008JCli...21.3521K,2022JCli...35.2633S}). For our purposes, we take the two climates of interest to be that obtained in a FRZ experiment (true effect of sea-ice loss), and the corresponding nudging or albedo modification experiment. 

The spurious response of non-atmospheric energy transport is due to nudging, as in our idealised model, ocean heat transport does not change between experiments. Likewise, the spurious response of top of atmosphere shortwave is determined solely by the upward shortwave flux, as the downward flux does not change between in experiments. Furthermore, in the idealised model, only changes in the surface albedo can affect the upward shortwave flux, which is the same at the surface and the top of atmosphere. The FRZ experiments capture the true response of surface albedo to sea-ice loss, so any differences between the albedo in NDG or ALB experiments, and FRZ experiments, are spurious. Taken together, these observations allow us to re-write Equation \ref{eq:I_eq} as an equation for the spurious response of the storm track intensity, $\Delta I^{\ast}$, to spurious energy input associated with the nudging and albedo modification methodologies: \begin{equation}
    \Delta I^{\ast} = \Delta I^{\ast}_{\text{LW}} - \Delta I^{\ast}_{\text{M}} + \Delta I^{\ast}_{\text{input}}, \label{eq:I_eq2a}
\end{equation}
with \begin{equation}
\Delta I^{\ast}_{\text{input}} = 2\pi a^{2} \int^{\vartheta}_{\frac{\pi}{2}} \left(F_{\text{nudge}} -\Delta\alpha_{\text{sfc}} F^{\downarrow}_{\text{sw, sfc}}\right)\cos\vartheta\text{d}\theta,
\end{equation}
which are Equations \ref{eq:I_eq2} and \ref{eq:Ispur} in the main text, respectively. Above, the asterisks denote a spurious forcing or response, and all differences are computed between experiments with spurious additional heating to melt ice (i.e., NDG, ALB) and the FRZ experiments, which do not introduce additional heating.

\bibliographystyle{ametsocV6}
\bibliography{manuscript}

\end{document}